\pdfoutput=1
\documentclass[12pt,english]{aastex}
\usepackage[latin9]{inputenc}
\usepackage{geometry}
\usepackage{amsbsy}
\usepackage{graphicx}
\usepackage{esint}

\makeatletter

\usepackage{amsbsy}

\makeatother

\usepackage{babel}
\begin{document}

\title{Magnetic helicity of global field in cycles 23 and 24}

\author{V.V. Pipin$^{1,3}$, A.A. Pevtsov$^{2}$}

\affil{ $^{1}$Institute of Solar-Terrestrial Physics, Russian Academy of
Sciences, Irkutsk, 664033, Russia \\
 $^{2}$ National Solar Observatory, Sunspot, NM 88349, USA \\
 $^{3}$ National Astronomical Observatories, Chinese Academy of Sciences,
Beijing 100012, China }
\begin{abstract}
For the first time we reconstruct the magnetic helicity density
of global axisymmetric field of the Sun { using} method proposed by 
\citet{2003AdSpR..32.1835B} and \citet{pip13M}. To determine the
components of the vector potential, we apply the gauge which is typically
employed in mean-field dynamo models. This allows for a direct comparison 
{ of reconstructed helicity} with the predictions from the mean-field dynamo
models. We apply the method to two different data sets: 
the synoptic maps of line-of-sight (LOS) magnetic field from the Michelson
Doppler Imager (MDI) on board of Solar and Heliospheric Observatory
(SOHO) and vector magnetic field measurements from Vector Spectromagnetograph
(VSM) on Synoptic Optical Long-term Investigations of the Sun
(SOLIS) system. Based on the analysis of MDI/SOHO data, we find that in solar cycle 23 
the global magnetic field had positive (negative) magnetic helicity in the 
northern (southern) hemisphere. This hemispheric sign asymmetry is opposite
to helicity of solar active regions, but it is in agreement with the predictions
of mean-field dynamo models. { The data also suggest that the
  hemispheric helicity rule may have 
 reversed its sign in early and late phases of cycle 23.}  
Furthermore, the data indicate an imbalance in magnetic helicity 
between the northern and southern hemispheres. This imbalance
seem to correlate with the total level of activity in each hemisphere in cycle 23.
Magnetic helicity for rising phase of cycle 24 is derived from SOLIS/VSM data,
and qualitatively, its latitudinal pattern is similar to the pattern derived 
from SOHO/MDI data for cycle 23.
\end{abstract}

\section{Introduction}

The generation of the magnetic field in the Sun is tightly related
with the convective helical motions. In the framework of axisymmetric
dynamos, the magnetic field is typically decomposed into toroidal
and poloidal components. 
\citet{P55} suggested that solar dynamo can be
represented as a periodic transformation of poloidal magnetic field,
$\bar{\mathbf{B}}^{(p)}={\bar{B}}_{r}\mathbf{e}_{r}+{\bar{B}}_{\theta}\mathbf{e}_{\theta}$,
into toroidal field $\bar{\mathbf{B}}^{(t)}={\bar{B}}_{\phi}\mathbf{e}_{\phi}$
(via the differential rotation) and the reverse transformation
of $\bar{\mathbf{B}}^{(t)}$ to $\bar{\mathbf{B}}^{(p)}$ by helical
convective motions. Further development of dynamo theory showed that the 
two processes produce 
helical magnetic fields on both small and large spatial scales 
\citep{pouquet-al:1975a,pouquet-al:1975b}, and that 
the conservation of magnetic helicity
is an important factor for the dynamical quenching of large-scale
magnetic field generation \citep{kleruz82,1991ApJ...376L..21C,1992ApJ...393..165V,kle-rog99,2000A&A...361L...5K,2005PhR...417....1B}.

Early observations of various proxies of magnetic/current helicity
established what is now known as the hemispheric helicity rule: 
magnetic fields of active regions exhibit preferentially negative (positive) helicity
in the northern (southern) hemisphere \citep[][and references therein]{see1990SoPh,pev95,zetal10}.

On the other hand, some researchers \citep[e.g.,][]{2003AdSpR..32.1835B,bl-br2003,warn2011,pip2013ApJ,pip13M} 
argued that magnetic helicity of large-scale (global) axisymmetric field should be positive/negative
in the northern/southern hemisphere. Furthermore, the mean-field dynamo models predict
reversals of the sign of helicity in association with the propagation of dynamo
wave inside the convection zone \citep[e.g.,][]{warn2011,pip13M}. Reversals of the hemispheric helicity rule 
have been reported in observations, but the results seems inconclusive,
with some researchers reporting the presence of such reversals 
in early/late phases of solar cycle \citep{bao2000,hag05}, while others are 
questioning their existence  \citep{pev2001,pev2008,2013ApJ...772...52G}.
{Such apparent controversy may be resolved via direct comparison of observations with 
the model predictions. Since dynamo models can
provide a detailed information about the distribution of magnetic helicity of global 
axisymmetric field in the convection zone and near the photosphere, it is highly 
desirable to directly compare these model estimates with the observations.}

Theory suggests that magnetic helicity on small and large scales should have
opposite sign \citep[e.g.,][]{pouquet76,seehafer96,bl-br2003,2005PhR...417....1B,pip2013ApJ,pip13M}. 
Then, the small-scale helicity may dissipate on small spatial
scales subject to the Ohmic dissipation \citep[e.g.,][]{pouquet76}
or the helicity of both signs could emerge through the solar photosphere.
Early measurements of vertical component of large-scale current helicity density,
\citep[e.g.,][]{2000ApJ...528..999P,2010ApJ...720..632W} found that
in its sign, the large-scale magnetic fields follow the same hemispheric
helicity rule as the active regions. These early studies concentrated
on spatial scales larger than the active regions but smaller then
the solar hemisphere. One should note that in the framework of mean-field dynamo, 
the magnetic fields of active regions represent the ``small-scale'' fields, while 
the large-scale fields refer to spatial scales comparable with the size
of solar hemisphere. In this article, we address the helicity
determination for large-scale magnetic fields as defined by mean-field dynamo
theory.  To avoid confusion with previous studies, we use the terms ``global'' and 
``large-scale'' to refer to magnetic fields on spatial scales much larger then active regions. 
We reconstruct the magnetic helicity of global axisymmetric
field using the approach suggested
by \citet{2003AdSpR..32.1835B} and  \citet[][see Section 2 below]{pip13M}. 
Section 3 describes the data sets 
and the reduction procedure. Section 4 presents our main results, and in Section 5 we 
discuss our findings.

\section{The formalism behind the computation of helicity of global axisymmetric
field}

Let us represent the axisymmetric magnetic field $\mathbf{\bar{B}}$
as 
\begin{eqnarray}
\bar{\mathbf{B}} & = & \mathbf{e_{\phi}}\bar{B}_{\phi}+\nabla\times\left(\bar{A}_{\phi}\mathbf{e_{\phi}}\right)=\nabla\times\bar{\mathbf{A}},\label{eq:bm}\\
\bar{\mathbf{A}} & = & \mathbf{r}\bar{A}_{r}+\bar{A}_{\phi}\mathbf{e_{\phi}}=\mathbf{r}T+\nabla\times\left(\mathbf{r}S\right),\label{eq:bmS}
\end{eqnarray}
where, $\mathbf{r}=r\mathbf{e}_{r}$, $\nabla\times\left(\mathbf{r}\bar{A}_{r}\right)=\mathbf{e_{\phi}}\bar{B}_{\phi}$.
Representation of the vector potential by equation (\ref{eq:bmS}) is
often employed in mean-field dynamo models. In the spherical coordinates,
the scalar functions $S$ (poloidal potential) and $T$ (toroidal potential), which are  
functions of $t$ (time), $r$ (radius), $\theta$ (polar angle) and $\phi$ (azimuth), 
are uniquely determined with the gauge \citep[e.g.,][]{KR80,ruz04}:
\begin{equation}
\int_{0}^{2\pi}\int_{-1}^{1}Sd\mu d\phi=\int_{0}^{2\pi}\int_{-1}^{1}Td\mu d\phi=0,\label{eq:norm}
\end{equation}
where $\mu=\cos\theta$. Equation (\ref{eq:norm})
is time-dependent, and it is applicable to arbitrary $r$ including the solar surface 
at $r=R_\Sun=R$. The magnetic helicity density is given 
by $\bar{\mathbf{A}}\cdot\bar{\mathbf{B}}$.
In this study we concentrate on the axisymmetric magnetic field, and thus, 
ignore a dependence of magnetic field components on azimuth $\phi$.

Let us suppose that we
have information about the axisymmetric components of the toroidal
field, $\bar{B}_{\phi}$ and the poloidal field, $\bar{B}_{r}$ from the observations. 
Then, decomposing $r$ and $\phi$ components of magnetic field and
its vector potential on series of the Legendre polynomial $P_{n}$ and $P_{n}^{1}$:
\begin{equation}
\bar{A}_{\phi}\left(t,\theta\right)=\sum_{n=1}^{N}a_{\phi}^{(n)}\left(t\right)P_{n}^{1}\left(\cos\theta\right),\label{eq:aa}
\end{equation}
\begin{equation}
\bar{B}_{r}\left(t,\theta\right)=\sum_{n=1}^{N}b_{r}^{(n)}\left(t\right)P_{n}\left(\cos\theta\right),\label{eq:br}
\end{equation}
\begin{equation}
\bar{B}_{\phi}\left(t,\theta\right)=\sum_{n=1}^{N}b_{\phi}^{(n)}\left(t\right)P_{n}^{1}\left(\cos\theta\right),\label{eq:bf}
\end{equation}
\begin{equation}
\bar{A}_{r}\left(t,\theta\right)=\sum_{n=1}^{N}a_{r}^{(n)}\left(t\right)P_{n}\left(\cos\theta\right),\label{eq:arf}
\end{equation}
and using a known relation between $P_{n}$ and $P_{n}^{1}$,
\[
P_{n}^{1}=-\sin\theta\frac{\partial P_{n}}{\partial\mu},\,-\frac{\partial P_{n}^{1}}{\partial\mu}=n\left(n+1\right)P_{n},
\]
one can derive the following algebraic relations between the coefficients for magnetic field and the vector potential Legendre polynomial series: 
\begin{eqnarray}
a_{\phi}^{(n)}\left(t\right) & = & -\frac{Rb_{r}^{(n)}\left(t\right)}{n\left(n+1\right)},\label{eq:nn}\\
a_{r}^{(n)}\left(t\right) & = & -Rb_{\phi}^{(n)}\left(t\right),\label{eq:nnb}
\end{eqnarray}
where $R$ is the radius of the Sun and $n$ is some finite (reasonably large) number. 
In the study we adopt
n$\ge48$. 

{The derivations of the vector potential and helicity in} \citet{2003AdSpR..32.1835B} 
are based on odd modes of equation (\ref{eq:nn}).
Here, we take into account both the radial and the toroidal magnetic fields, and
we use information from both even and odd modes of the coefficients in equations 
(\ref{eq:aa}-\ref{eq:arf}).
This approach allows us to study both symmetric (relative to the solar equator) and antisymmetric 
components of magnetic helicity density. To calculate a proxy for poloidal component of 
large-scale current helicity density, $\bar{B}_{r}\left(\nabla\times\mathbf{\bar{B}}\right)_{r}$,
we employ the following identities: 
\begin{eqnarray}
\left(\nabla\times\bar{\mathbf{B}}\right)_{r}\left(t,\theta\right) & = & \sum_{n=1}^{N}\left(\nabla\times\bar{\mathbf{B}}\right)_{r}^{(n)}\left(t\right)P_{n}\left(\cos\theta\right),\label{eq:curr}\\
\left(\nabla\times\bar{\mathbf{B}}\right)_{r}^{(n)}\left(t\right) & = & -\frac{n\left(n+1\right)b_{\phi}^{(n)}\left(t\right)}{R}
\end{eqnarray}
Scalar function $S$ (poloidal potential, see equation \ref{eq:bmS}) can be determined 
within the uncertainty of a gauge (constant), which does not affect the vector potential of poloidal field (toroidal part of  the vector potential $\mathcal{A}_{\phi}$).  The gauge in equation (\ref{eq:norm}) only affects the vector potential of the toroidal field (poloidal component of vector potential $\mathcal{A}_r$).
This component of vectors potential is determined via equations (\ref{eq:arf},\ref{eq:nnb}).
In the reconstruction of vector potential,
equation(\ref{eq:norm}) can be satisfied numerically by re-defining
$\bar{A}_{r}(t,\mu)=\bar{A}_{r}^{(0)}(t,\mu)-\pi C(t)$, where 
$\bar{A}_{r}^{(0)}\left(t,\mu\right)={\displaystyle \sum_{n=1}^{N}a_{r}^{(n)}\left(t\right)P_{n}\left(\mu\right)}$
and $C$ can be determined numerically from the integration: 
\begin{equation}
\int_{0}^{2\pi}\int_{-1}^{1}\bar{A}_{r}^{(0)}\left(t,\mu\right)d\mu d\phi=2\pi\int_{-1}^{1}\bar{A}_{r}^{(0)}\left(t,\mu\right)d\mu=C\left(t\right),\label{eq:tnn}
\end{equation}
In the course of the reconstruction, we found that the amplitude of $C\left(t\right)$
is rather small in comparison with $\bar{A}_{r}^{(0)}(t,\mu)$. 
Equations (\ref{eq:bm}--\ref{eq:norm},\ref{eq:tnn}) ensure that $\int_{-1}^{1}\bar{A}_{\phi}\bar{B}_{\phi}d\mu=\int_{-1}^{1}\bar{A}_{r}\bar{B}_{r}d\mu$
which should be expected from the topological considerations \citep[see Section 4 and][]{2003AdSpR..32.1835B}. Additional details about the formalism employed in computation of helicity
of global field can be found in \citet{pip13M}. 

\section{Data reduction}

\subsection{The construction of maps }

\begin{figure}
a)\includegraphics[width=0.45\columnwidth]{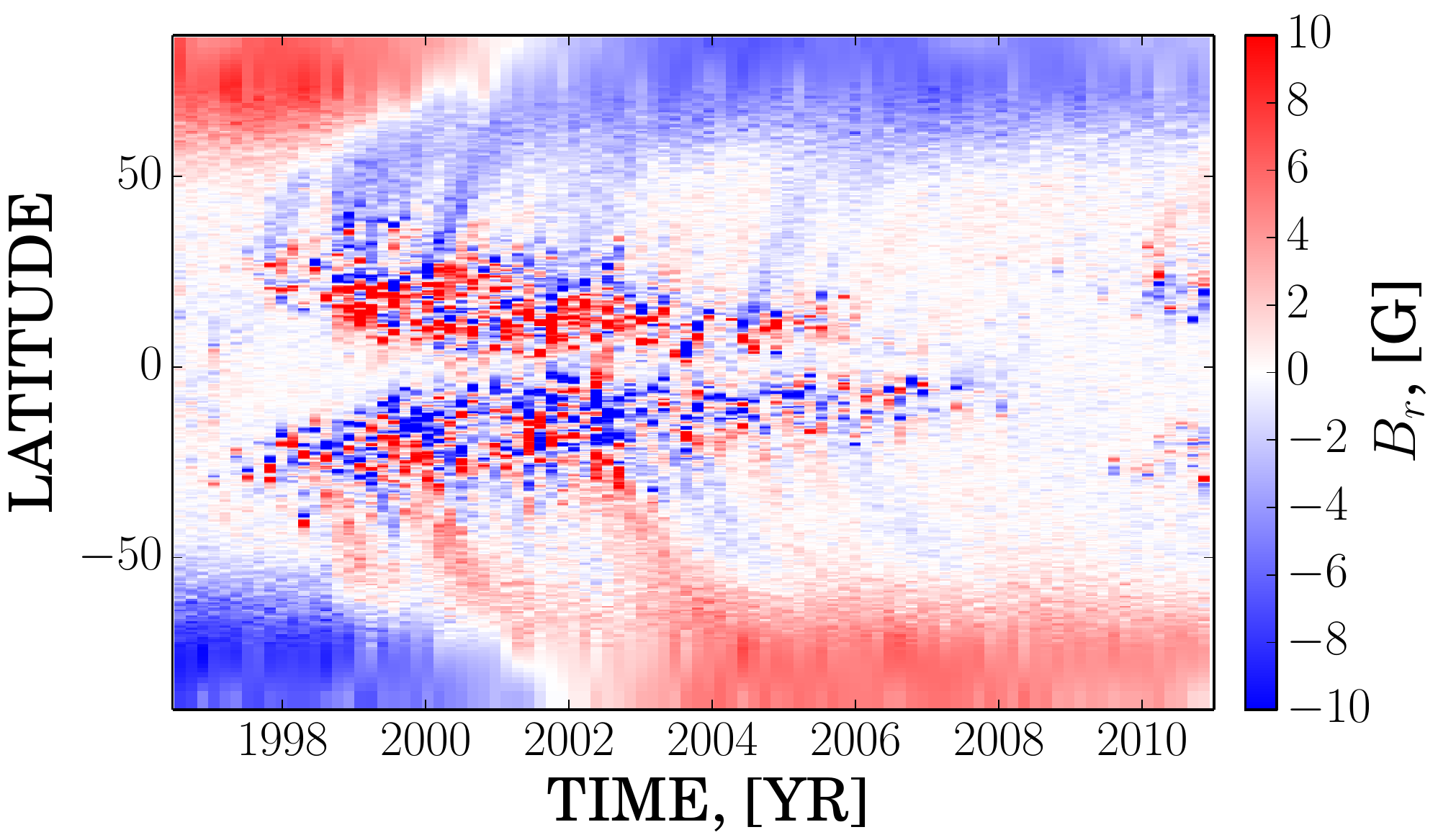}

b)\includegraphics[width=0.45\columnwidth]{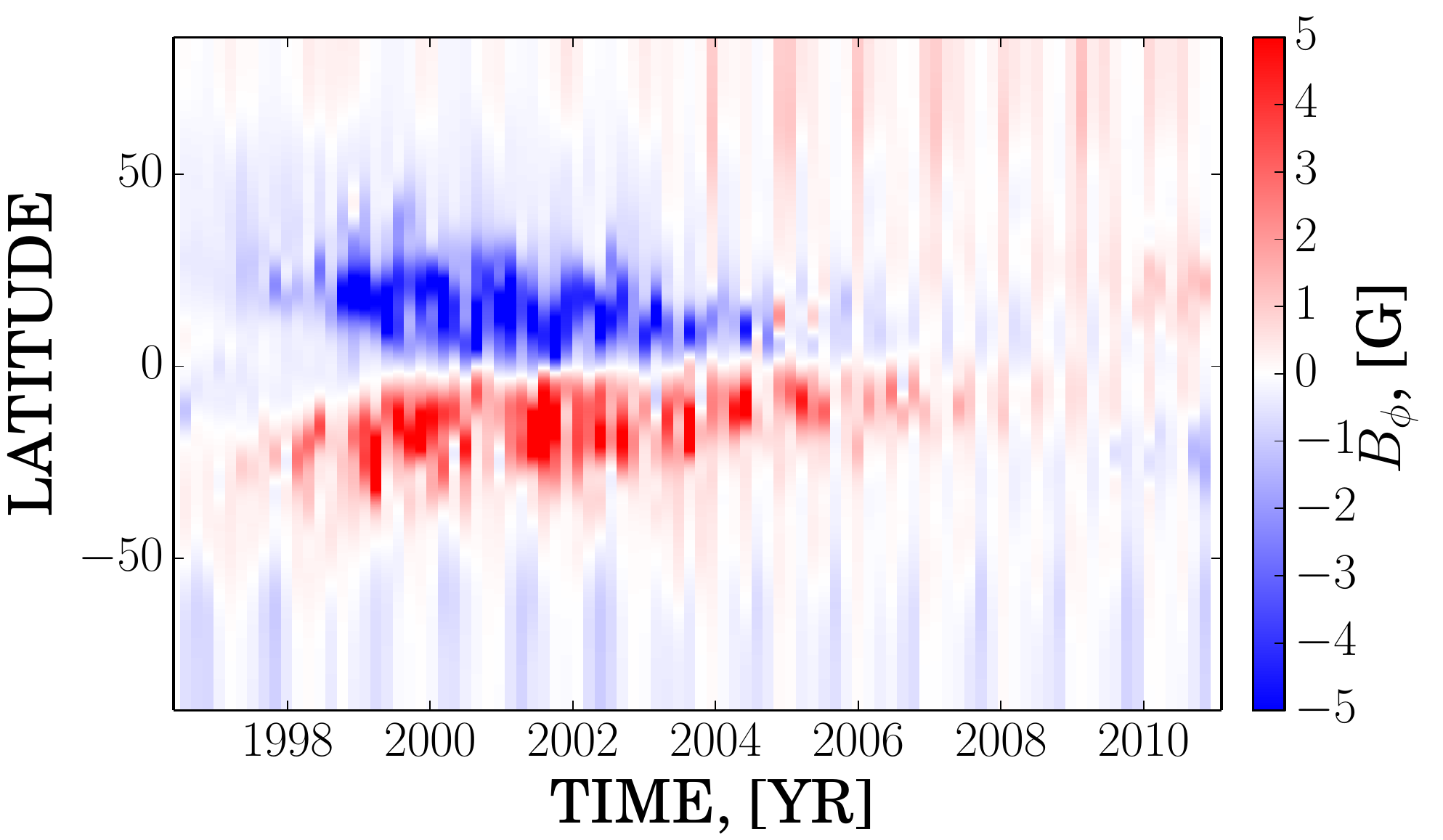}

c)\includegraphics[width=0.45\columnwidth]{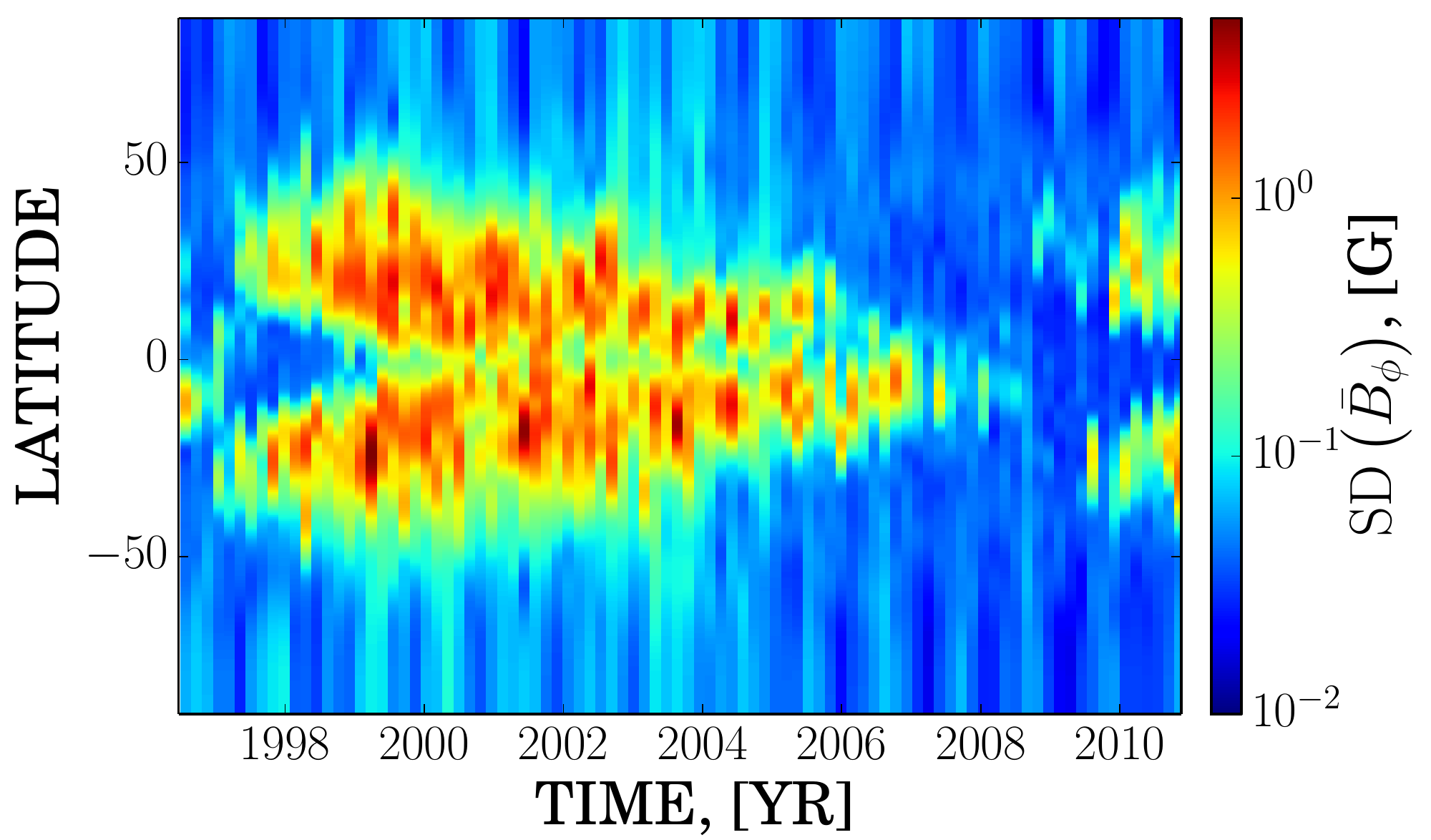}

\caption{\label{fig:LOS}The time-latitude evolution of the large-scale magnetic
field derived from MDI synoptic charts: a) the $\bar{B}_{r}$ component; b) the $\bar{B}_{\phi}$ component;
the standard error of the $\bar{B}_{\phi}$. Vertical stripes in the
toroidal flux and its error are the results of annual variations in
the latitude of solar disk center (so called B$_{0}$--angle).}
\end{figure}

Next, we employed the synoptic maps of line-of sight (LOS) magnetic fields from the Solar and Heliospheric
Observatory/Michelson Doppler Imager (SOHO/MDI) data set \citep{1995SoPh..162..129S,2004SoPh..219...39L,2010AGUFMSH41D..02H,2011SoPh..270....9S}.
The $\bar{B}_{r}$ and $\bar{B}_{\phi}$ components of magnetic field are
derived from a set of synoptic maps constructed using 10-degree wide longitudinal 
segments of solar disk image centred at the following longitudes relative to
the central meridian: $\phi_{i}=0,\pm(15,30,45$$^{\circ}$ and $60$$^{\circ})$.

The poloidal and toroidal components of magnetic field can be determined following 
\citet{WSO1} approach, under the assumption that the observed changes in the magnetic 
fields over several days are entirely due to the change in the projection of the same
magnetic field vector. Ideally, the
method requires to compare the same feature on all images (see, \citealp{WSO1,grig86,2000ApJ...528..999P,2010ApJ...720..632W}). Instead, we employ a more 
simplified approach by comparing areas with the same latitudes and longitudes
in synoptic maps constructed for different longitudinal offsets.

For each pixel on these synoptic maps of LOS field
\begin{equation}
B_{l}\left(t,\phi_{i},\theta\right)=\bar{B}'_{r}\left(t,\theta\right)\sin\theta\cos\phi_{i}+\bar{B}_{\phi}\left(t,\theta\right)\sin\phi_{i},\label{eq:torf}
\end{equation}
{where $B_{l}$ is the line-of-sight component of magnetic
field and $\bar{B}'_{r}$ and $\bar{B}_\phi$ are its radial (poloidal) and
toroidal component of global magnetic field (here we introduce $\bar{B}'_{r}$ 
notation to distinguish from $\bar{B}_{r}$ computed by SOHO/MDI team using a 
different approach). One can see that a simple addition/subtraction of synoptic 
maps taken with longitudinal offsets symmetric relative to the central meridian (e.g., $\pm$ 30$^\circ$)
allows to determine $\bar{B}'_{r}$ and $\bar{B}_\phi$ from the equation \ref{eq:torf}.
To lessen the effects of magnetic field evolution, we smooth each synoptic map by convolving it
with a symmetric 2D Gaussian function with the FWHM of four solar degrees. 
Next, we determine the toroidal and poloidal components
by fitting equation (\ref{eq:torf}) for each Carrington longitude using
the data taken with different longitudinal offsets. Finally, we average
the obtained components of the magnetic field vector over each Carrington
rotation to derive the latitudinal profiles of $\bar{B}'_{r}$ and $\bar{B}_{\phi}$.
As a test, we compared the $\bar{B}'_{r}$ with $\bar{B}_{r}$ from the synoptic
charts of radial solar magnetic field provided by the MDI team \citep{2011SoPh..270....9S}.
We found that the latitudinal profiles of $\bar{B}'_{r}$ derived by us
agrees well with $\bar{B}_{r}$ profiles derived by SOHO/MDI team. 
Some minor deviations were found at very high latitudes near the polar regions. 
The latter could be explained by the fact that SOHO/MDI synoptic maps of radial ($\bar{B}_{r}$) 
field employ the polar field filling, while we employ $B_l$ synoptic charts without
pole-filling. In further computations we use the latitudinal profiles of $\bar{B}_{r}$.
As other test, for three solar rotations (CR1913, 1979, 2058), we compared
the latitudinal profiles of radial and toroidal fields derived
by us with those from \citet{2010ApJ...720..632W}. We found a good
agreement between two independent derivations. 

Figure \ref{fig:LOS} demonstrates the derived distribution of radial and toroidal
components of magnetic field for the entire SOHO/MDI data set. One can see
several well-known patterns. For example, at high latitudes, the radial
flux (Figure \ref{fig:LOS}a) shows weak polar fields of correct
sign, as well as the polar field reversals shortly after the maximum
of cycle 23. In mid-latitudes, the prevailing polarity field is positive/negative
in the northern/southern hemisphere, in agreement with the leading
polarity of active region fields for cycle 23. 
As the leading polarity flux in active regions is more compact 
(and hence, may last longer) in comparison with the following polarity flux, 
average synoptic charts such as Figure \ref{fig:LOS} tend to emphasize the leading fields.
The toroidal field
(Figure \ref{fig:LOS}b)in active region belts is negative/eastward
in the northern hemisphere (and it is positive/westward in the southern
hemisphere) in agreement with the prevailing polarity orientation
of active regions in cycle 23 (i.e., Hale polarity rule). The data
also show a weak (but persistent) pattern of east-west inclination of solar
magnetic field, which is in agreement with \citet{hoeks10} and \citet{2011SoPh..270....9S}
findings.

Figure \ref{fig:LOS}b shows that at high latitude (near polar) regions there is a weak
but persistent toroidal field oriented in opposite direction
to the field of active regions. Thus, for example, weak fields in the declining
phase of cycle 23 are oriented westward (positive)/eastward (negative)
in the northern/southern hemisphere. In combination with the polarity
of polar field in cycle 23, this implies that the weak fields outside
of active regions are inclined (pointing towards) up-eastward in the southern hemisphere
and down-westward in the northern hemisphere. Such tilt was previously
noted by \citet{WSO1,2000ApJ...528..999P}. One could also note, that
at high latitudes, the sign of the toroidal component of weak field 
corresponds to the
orientation of active region magnetic fields in the next cycle
24 as if these weak fields herald the cycle 24 starting at high latitudes
well before the first active region of this cycle emerges. \citet{tlatov10,tlatov13}
found signs of the extended solar cycle in the orientation of
ephemeral active regions several years prior to beginning of sunspot
cycle, which qualitatively agrees with high latitude patterns shown in Figure \ref{fig:LOS}.

\subsection{Mitigation of orbital periodicity and the reduction of noise}

Toroidal flux shows the effects related to a one year orbital periodicity
and the presence of a noise component. To mitigate the effects of 
orbital periodicity, we employ the following strategy. Since the toroidal
magnetic field and the toroidal vector potential should be zero at
the poles, we restricted the computation of $\bar{B}_{\phi}$ to $\pm$70$^\circ$
latitudes. For latitudes between 70$^\circ$ and 90$^\circ$ the
toroidal field was extrapolated linearly from lower latitudes.

Noise in toroidal flux was reduced by convolving the data (Figure \ref{fig:LOS})
with 2D Gaussian function: 
\begin{equation}
G(\mu,t)=\exp\left(-\frac{\mu^{2}}{2b^{2}}\right)\left[\exp\left(-\frac{t^{2}}{2a^{2}}\right)-e^{-2}\left(3-\frac{t^{2}}{2a^{2}}\right)\right],\label{eq:kern}
\end{equation}
where $t$ is a discrete time given in units of Carrington rotations (CR).
The data are defined at the homogeneous mesh in $\mu$. For
spatial $\mu$ coordinate we employ the filter with the full width at
half maximum (FWHM) to be equal to 20 points of mesh, which corresponds
to $b\approx0.04$ in equation \ref{eq:kern}. For time coordinate, the FWHM is equal to 24
CR, i.e, $a=12$ CR. In addition, we apply reflection conditions at
the boundaries for spatial component of filter and vanishing first derivative
at the end-points for the time component of the filter (second
term in square brackets in equation \ref{eq:kern}). Similar filter
is usually applied for sunspot number analysis \citep{hath09}.

\subsection{The derivation of vector potentials}

Figure \ref{data-smooth} shows the derived symmetric and asymmetric
(relative to equator) components of radial and toroidal magnetic fields.
(One can note that the amplitude of solar cycle variations in symmetric
components is smaller as compared with the asymmetric one. We postpone
the discussion of this till Section 4.) Following \citet{zetal10},
the window sizes of temporal and spatial scales
involved in equation (\ref{eq:kern}) are chosen to correspond to
the scales of turbulent diffusion (we thanks K.M. Kuzanyan for
this idea), which is about $10^{12}$cm$^{2}/$sec
\citep{abr11}.

\begin{figure}
a)\includegraphics[width=0.49\columnwidth]{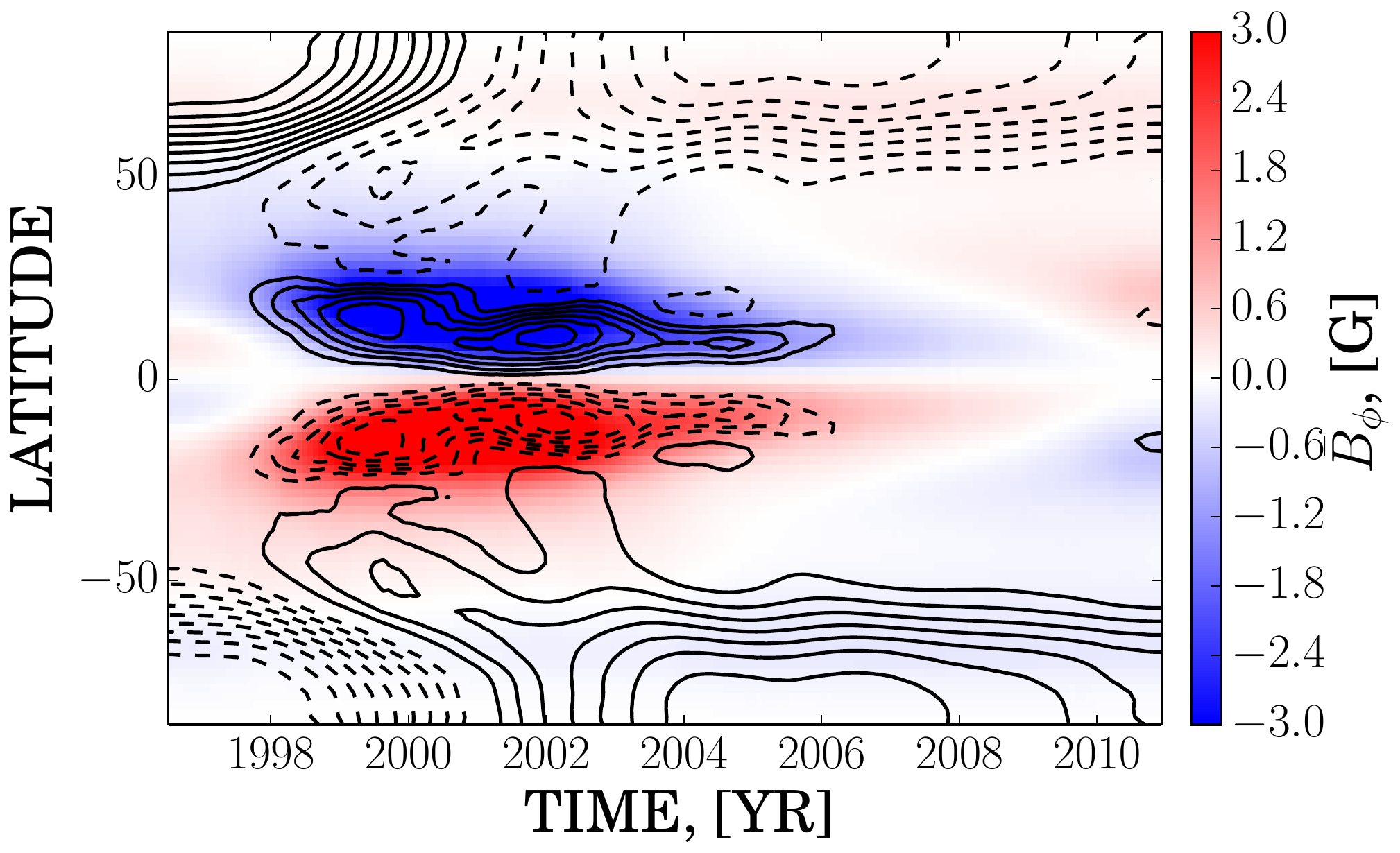}b)\includegraphics[width=0.49\columnwidth]{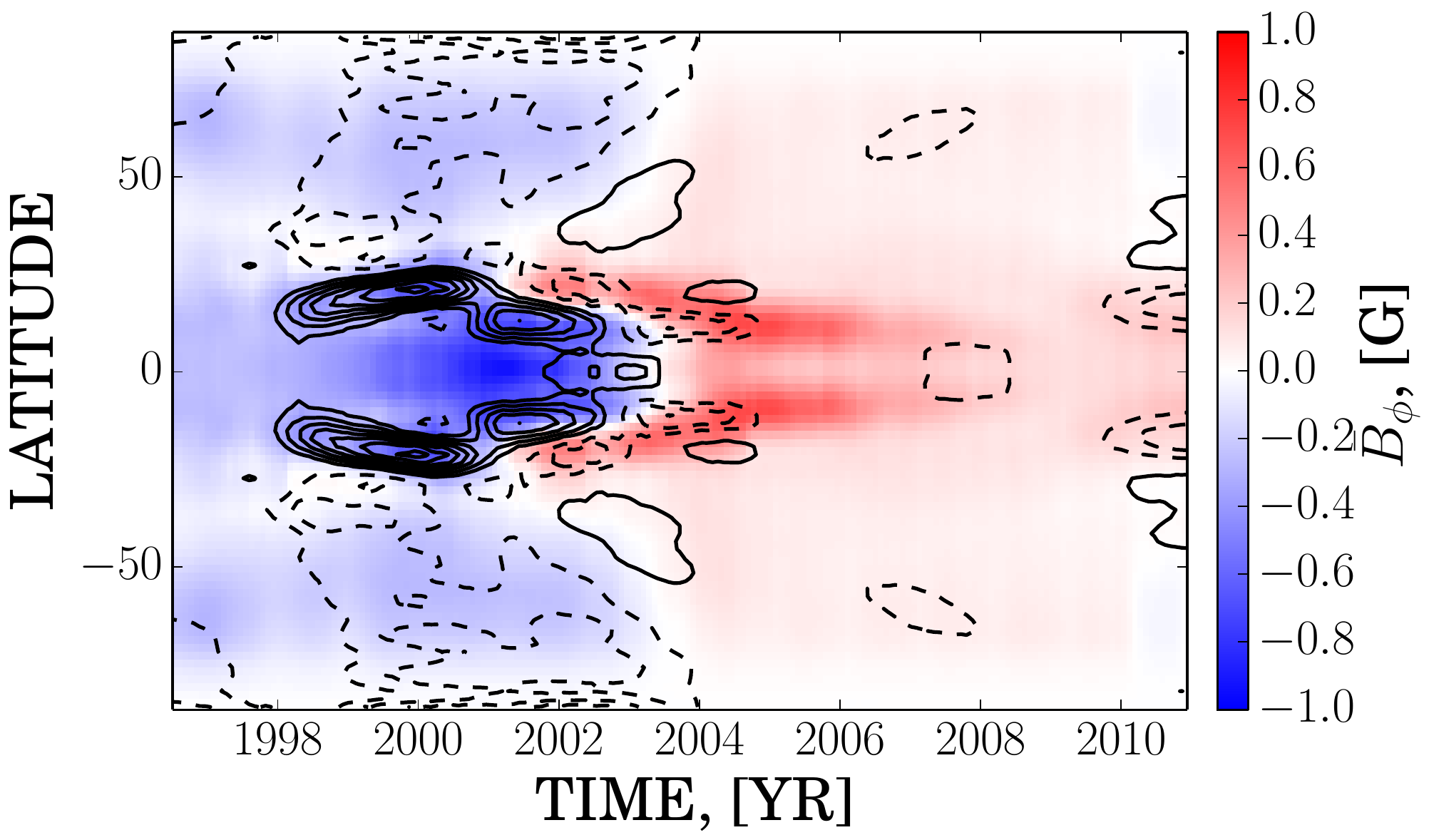}

\caption{\label{data-smooth}The time-latitude variations of the toroidal (background
image), $\bar{B}_{\phi}$, and the radial field, $\bar{B}_{r}$ (contours), after
convolution with kernel defined by equation (\ref{eq:kern}). Panels (a) and (b) show asymmetric
and symmetric components of magnetic field, accordingly. Positive (solid) and negative 
(dashed) contours are drawn at equal steps in amplitude of $\bar{B}_r$ within the range 
of $\pm6$G (panel a) and $\pm$3G.}
\end{figure}

Next, we interpolate the data to the collocation points of the Legendre
polynomials, $\mu_{j}=\cos\theta_{j}$, which are taken at zeros of
$P_{n}\left(\mu\right)$. The order of the polynomial approximation,
n, should be sufficiently high. We found that the results do not change
significantly for $n\ge48$, which was the basis for selecting N$=$48 as upper limit for
summation in equation \ref{eq:radial-ll}.
The coefficients $a_{\phi}^{(n)}(t)$ in equation(\ref{eq:aa}) can be
found using the equation(\ref{eq:bm}) and properties of the Legendre polynomials.
The matrix equation for $a_{\phi}^{(n)}\left(t\right)$ becomes 
\begin{eqnarray}
R\bar{B}_{r}\left(t,\mu_{j}\right) & = & -\sum_{n=1}^{N}a_{\phi}^{(n)}\left(t\right)n\left(n+1\right)P_{n}\left(\mu_{j}\right),\label{eq:radial-ll}\\
\bar{B}_{\phi}\left(t,\mu_{j}\right) & = & \sum_{n=1}^{N}b_{\phi}^{(n)}\left(t\right)P_{n}^{1}\left(\mu_{j}\right).
\end{eqnarray}
By solving the matrix equations in the collocation points, one can
find the coefficients for the vector potential components and restore
the distribution of magnetic helicity density. The validity of this
reconstruction procedure was tested using the output of a mean-field
dynamo model of \cite{pip13M}.

The main conclusions of this paper are drawn from the analysis based
on LOS magnetic field synoptic maps. These maps cover the solar
cycle 23 and the beginning of cycle 24. In addition, for the rising
phase of solar cycle 24 we employed the vector synoptic maps
from Vector Spectromagnetograph (VSM) on Synoptic Optical Long-term
Investigations of the Sun (SOLIS) system \citep{2013ApJ...772...52G}.
The maps cover 20 consecutive solar rotations starting from the
CR2109; this data set covers the period from March 2011
to December 2012. Time-latitude distribution of radial and toroidal 
components of vector magnetic field are shown in
Figure \ref{nso1}(a,b). Figure \ref{nso1}(c) shows the average latitudinal
profiles of two components of axisymmetric large-scale magnetic field. The profiles
were obtained by averaging over the Carrington rotations 2109--2128
and applying the Gaussian filter with FWHM equal 30 pixels in sine of latitude.

\begin{figure}
a)\includegraphics[width=0.45\columnwidth]{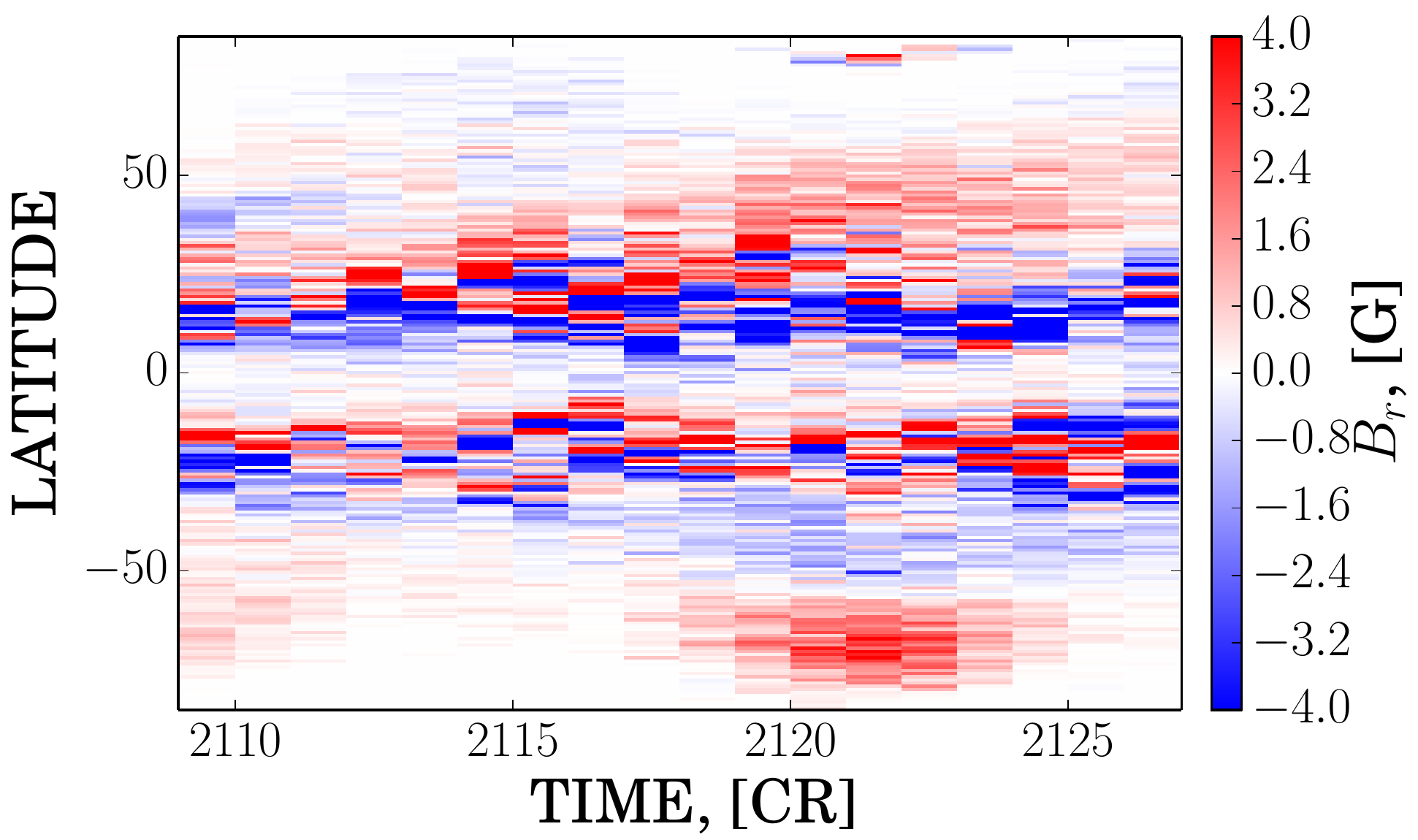}

b)\includegraphics[width=0.45\columnwidth]{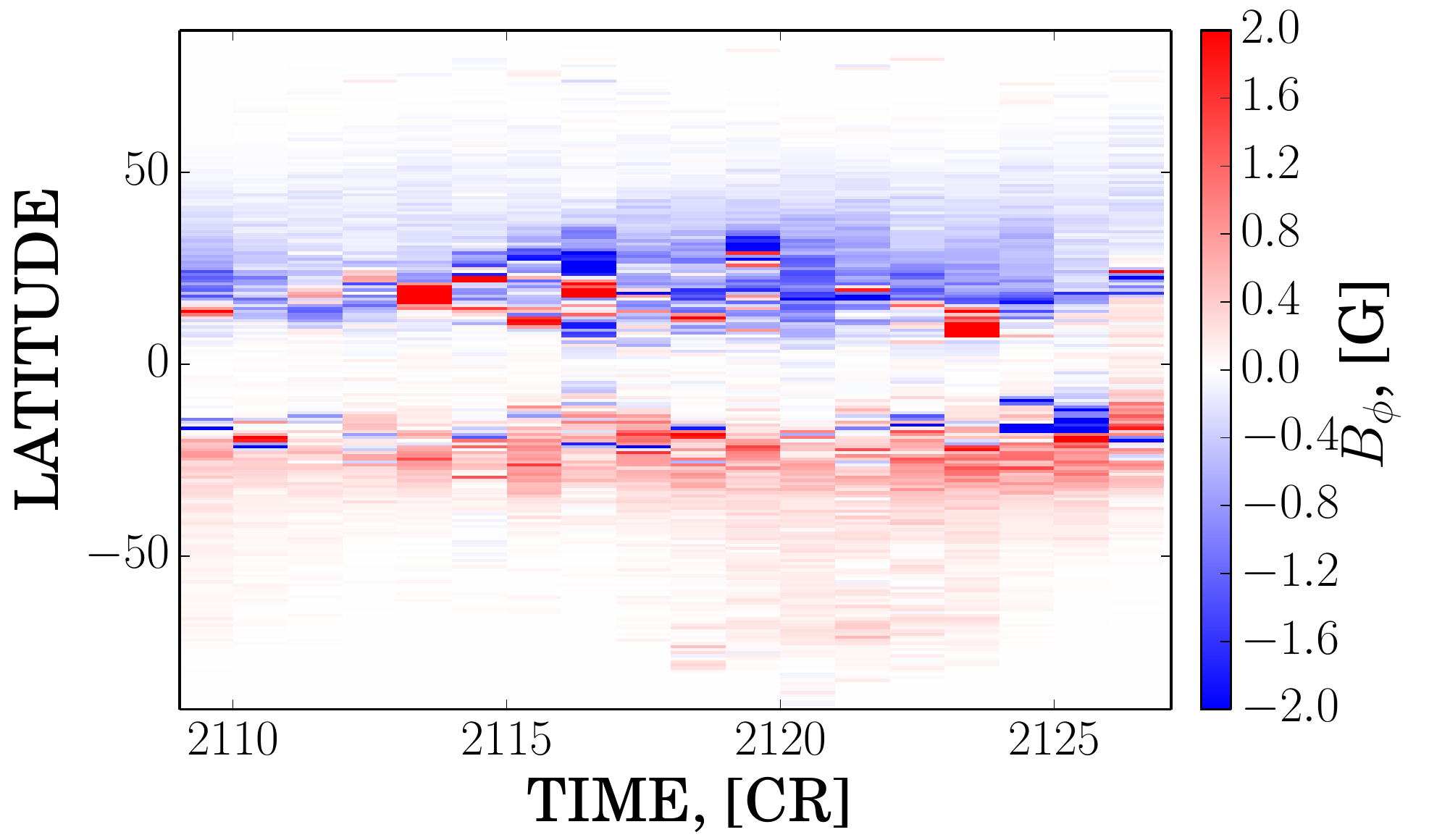}

c)\includegraphics[width=0.45\columnwidth]{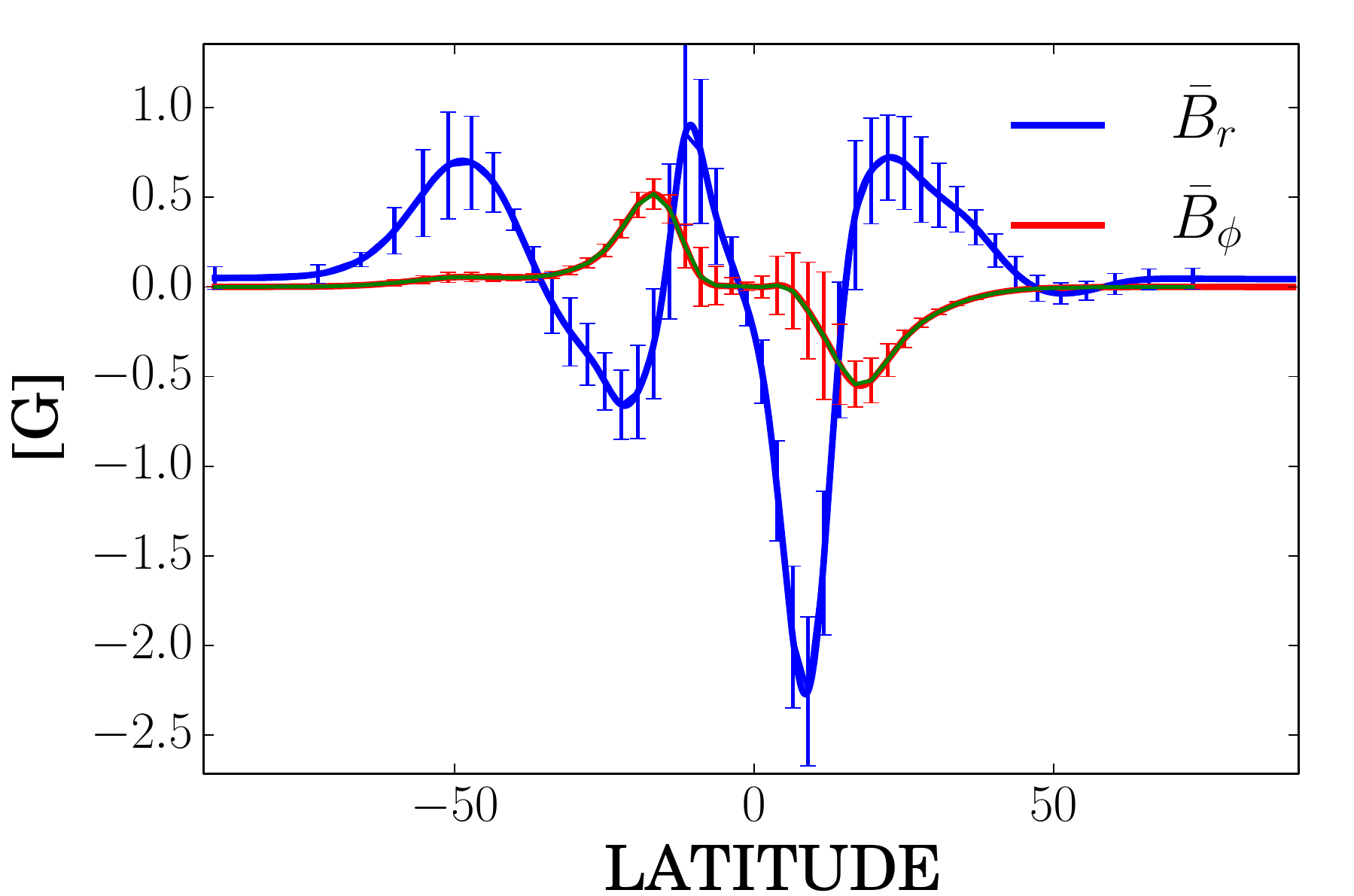}

\caption{\label{nso1}Radial ($\bar{B}_{r}$, panel a) and toroidal ($\bar{B}_{\phi}$, panel b) 
magnetic fields derived from SOLIS/VSM synoptic vector charts. Panel (c)
shows the mean latitudinal profiles of $\bar{B}_{r}$ and $\bar{B}_{\phi}$
together with the 90 \% confidence interval. }
\end{figure}

While there is no overlap between the MDI and VSM datasets to allow
for a more direct comparison, we note that the distributions of radial
and toroidal fields from two data sets exhibit somewhat similar behavior.
For example, similar to Figure \ref{fig:LOS} mean toroidal field
derived from the vector data is mostly negative in the northern hemisphere,
and it is mostly positive in the southern hemisphere. The polarity of radial field
in the main peak (negative in the northern hemisphere and positive
in the southern hemisphere) corresponds to the leading and following
polarity fields of dissipating active regions. MDI data (Figure \ref{fig:LOS}a)
show similar patterns in some parts of cycles 23 and 24 (e.g., see
``tip'' of cycle 24 ``butterfly'' in the northern hemisphere).
These general similarities provide some level of confidence for our method of derivation
of radial and toroidal components of large-scale magnetic field from
MDI synoptic maps of LOS flux.

\section{Results}

Figure \ref{modes2}(a,b) presents the evolution of power spectra
$\sqrt{\left[b_{\phi}^{(n)}\left(t\right)\right]^{2}}$ and $\sqrt{\left[a_{\phi}^{(n)}\left(t\right)\right]^{2}}$.
The coefficients $b_{\phi}^{(n)}\left(t\right)$ and
$a_{\phi}^{(n)}\left(t\right)$ decay rapidly with the increasing
number of modes. Furthermore, we found that the asymmetric component
of the magnetic field exhibits a faster decay, which we interpret
as if this component being more global in its nature as compared with
the symmetric component. In a hindsight, we note that one can draw a similar conclusion
using the results of \citet{sten88} study.

\begin{figure}
a)\includegraphics[width=0.49\columnwidth]{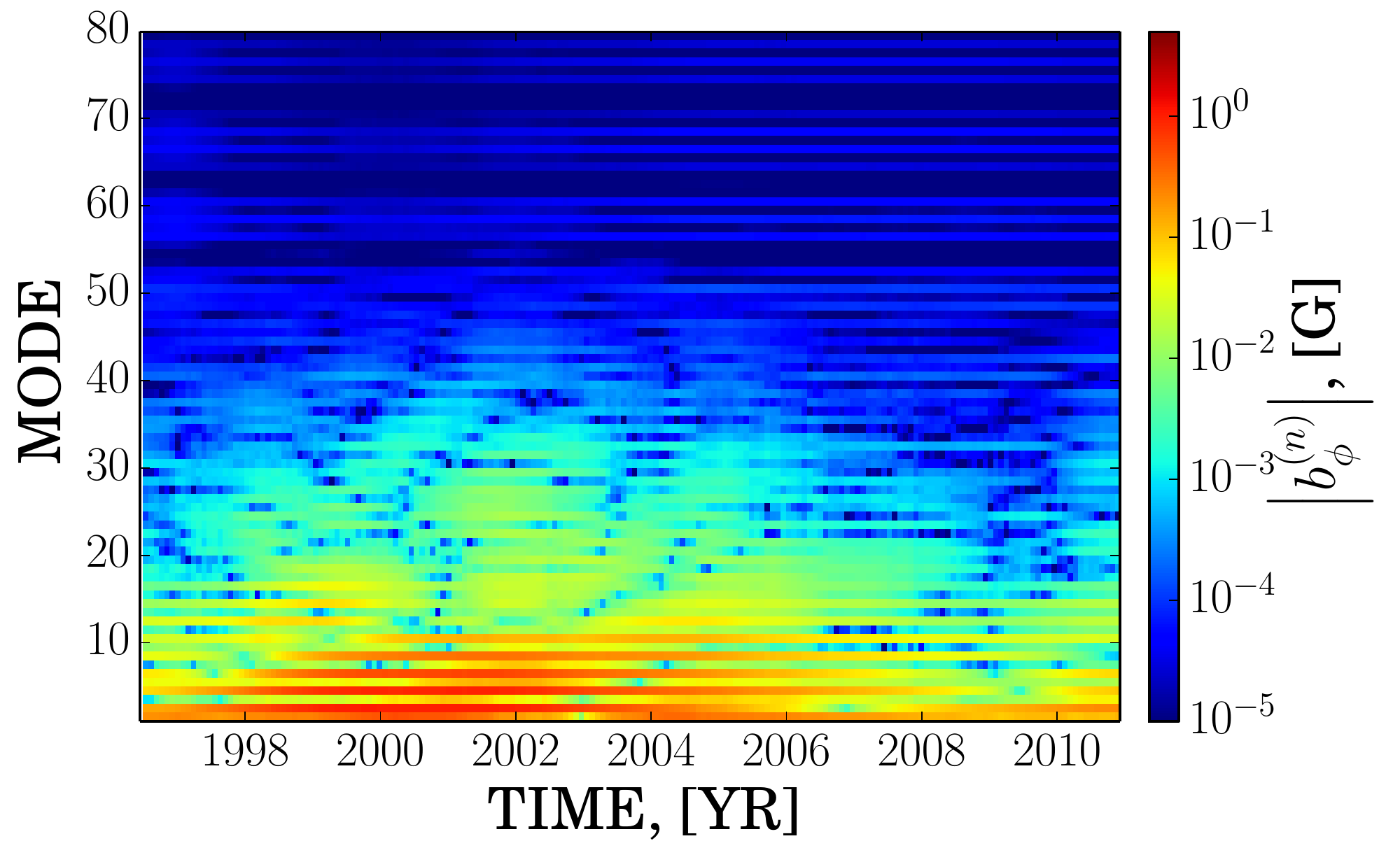}b)\includegraphics[width=0.49\columnwidth]{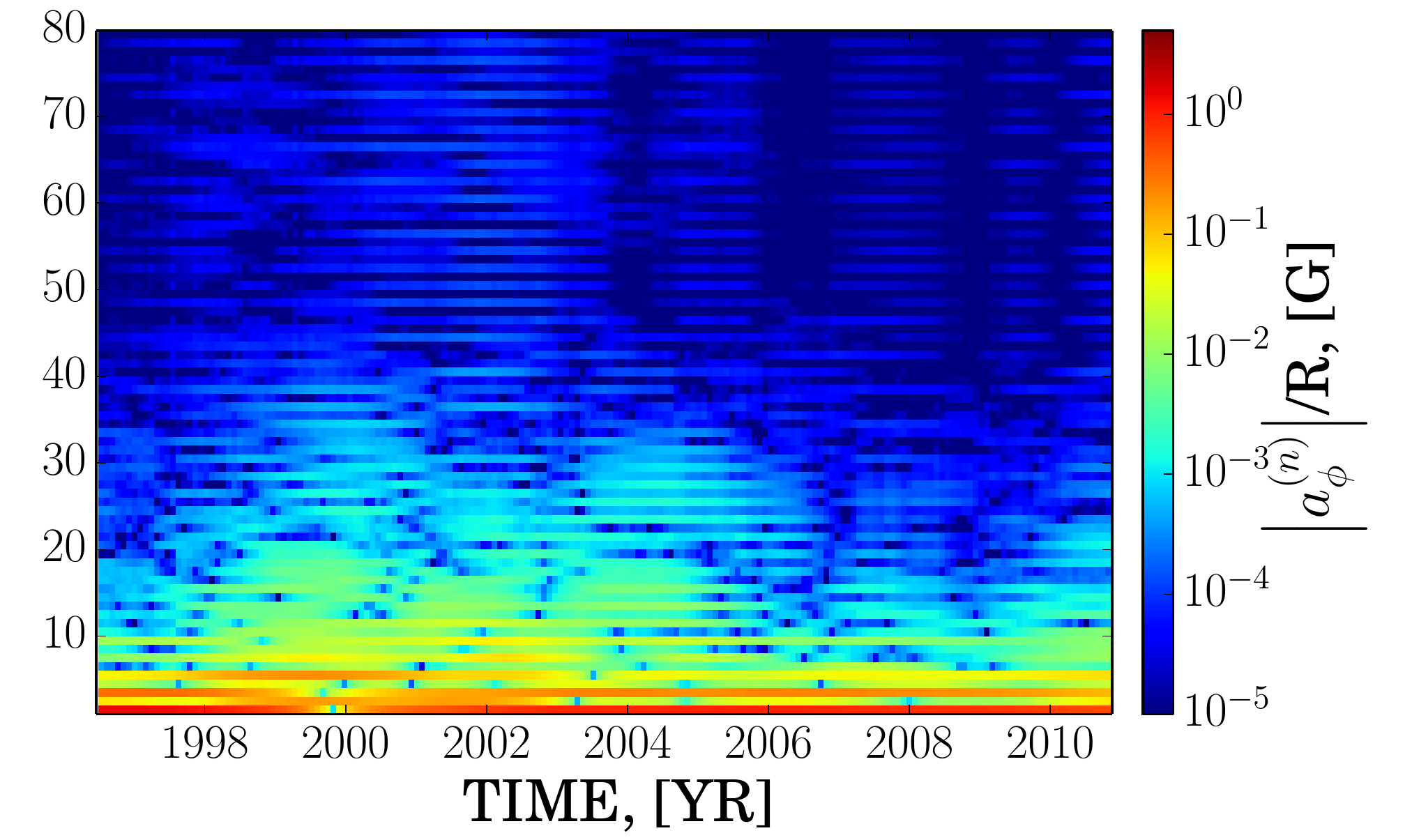}

\caption{\label{modes2}The evolution of power spectra, $\sqrt{\left[b_{\phi}^{(n)}\left(t\right)\right]^{2}}$
(panel a) and $\sqrt{\left[a_{\phi}^{(n)}\left(t\right)\right]^{2}}$ (panel b).}
\end{figure}

Figure \ref{fig:Components-of-the} shows the reconstructed components
of vector potential which are computed for two cases: (1) taking
into account odd and even modes of the spectral harmonics and
(2) including only the even modes (associated with the asymmetric
part of global magnetic field). There, we used the first
11 modes in equations (\ref{eq:aa},\ref{eq:arf}). Restricting
the expansion to 11 modes is well-justified by a rapid decay of $a^{(n)}\left(t\right)$
for higher modes (Figure \ref{modes2}b).

\begin{figure}
a)\includegraphics[width=0.49\columnwidth]{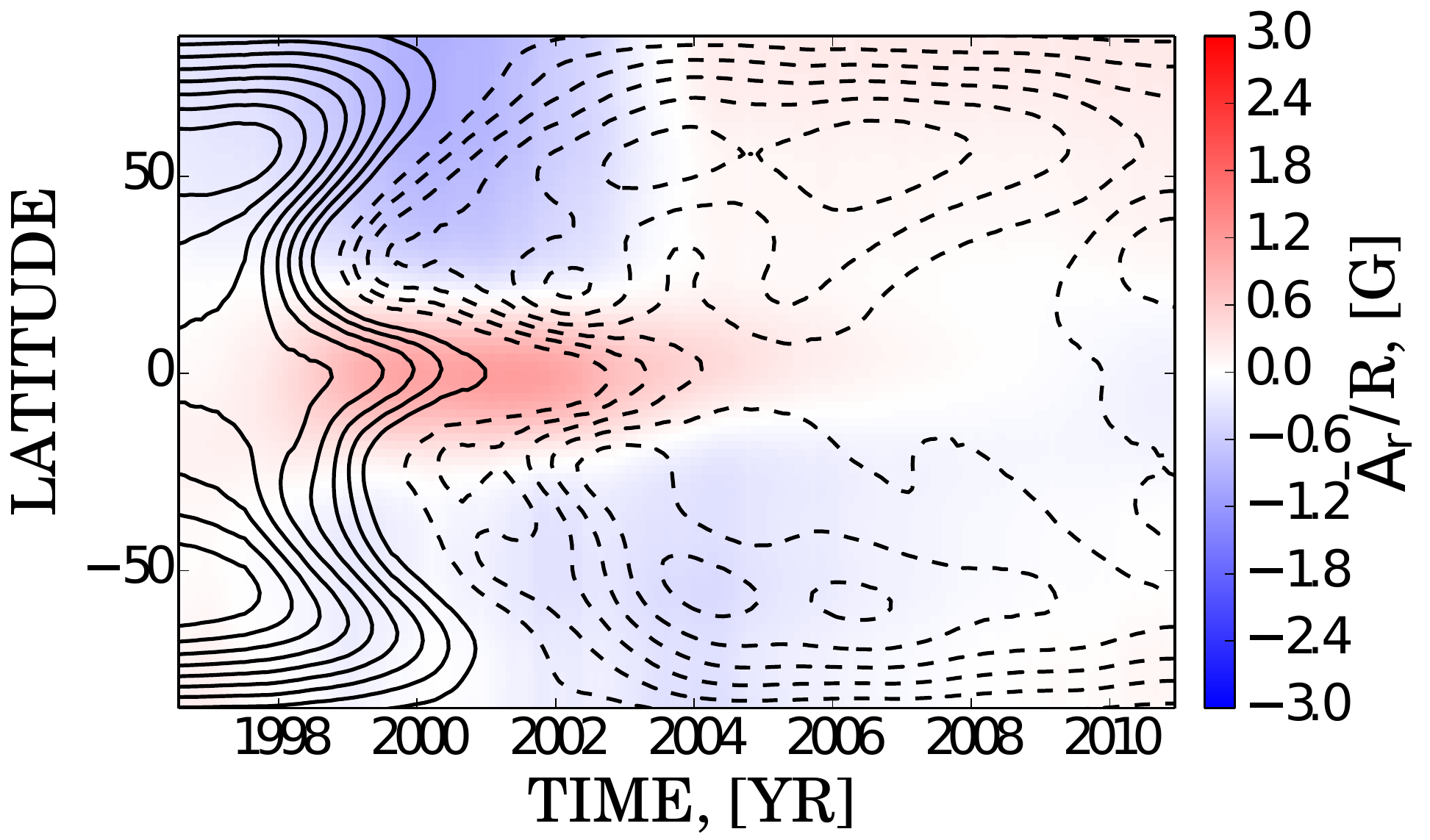}b)\includegraphics[width=0.49\columnwidth]{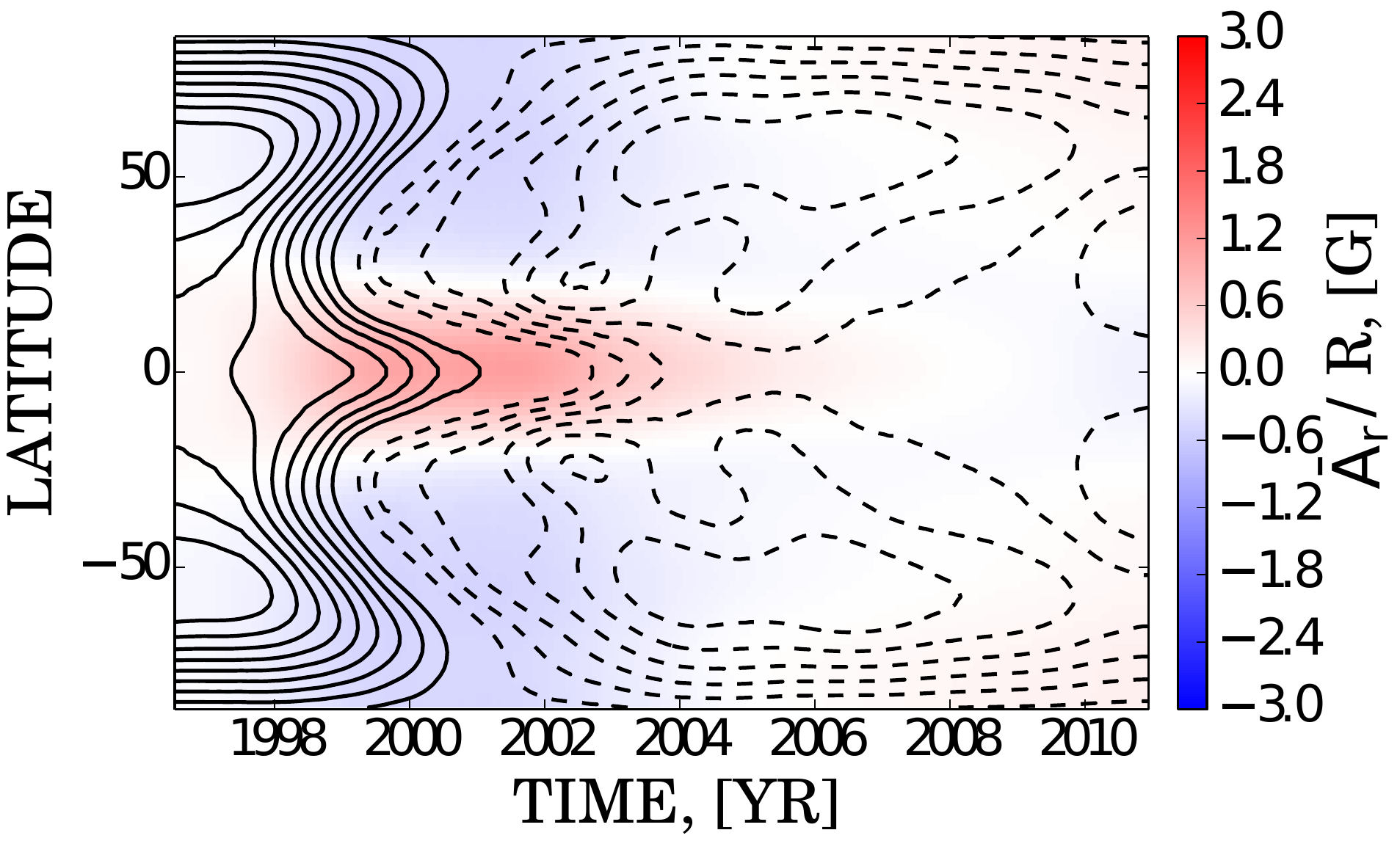}

\caption{\label{fig:Components-of-the}Radial ($\bar{A}_{r}/R$, background
images) and toroidal ($\bar{A}_{\phi}/R$, contours) components of
vector potential for (a) case 1 (including both odd and even modes)
and (b) case 2 (only even modes). Positive (solid) and negative (dashed) contours are drawn at equal steps in amplitude of $\bar{A}_{\phi}/R$ within the range of $\pm2$G.}
\end{figure}

The pattern of symmetric (relative to the equator) component of $\bar{A}_{\phi}$ is similar 
in appearance to reconstruction
made by \citet{2003AdSpR..32.1835B} based on \citet{sten88}
data. Note, that in our case we don't restrict the study to
a particular symmetry of the global field about the equator. We find that
the poloidal component of vector potential (Figure \ref{fig:Components-of-the}a)
exhibits a break in the equatorial symmetry (see change in sign of
$\bar{A}_{r}$ around year 2004). The asymmetry between northern and
southern hemispheres is also present at high latitudes prior to year
1998. On the other hand, the pattern of the $\bar{A}_{\phi}$ (Figure
\ref{fig:Components-of-the}b) exhibits no significant changes over
the solar cycle 23.

\begin{figure}
a)\includegraphics[width=0.47\columnwidth]{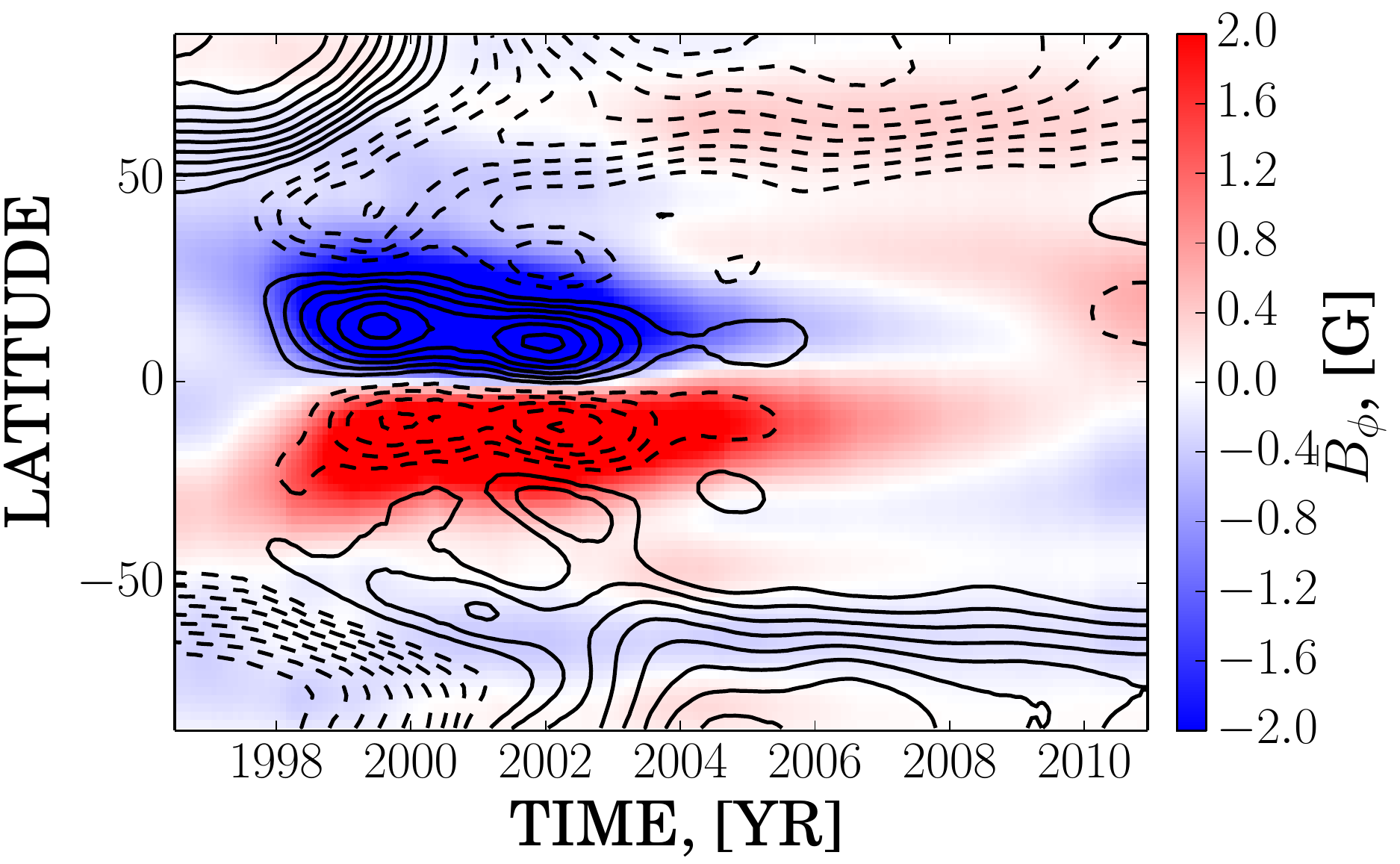}b)\includegraphics[width=0.47\columnwidth]{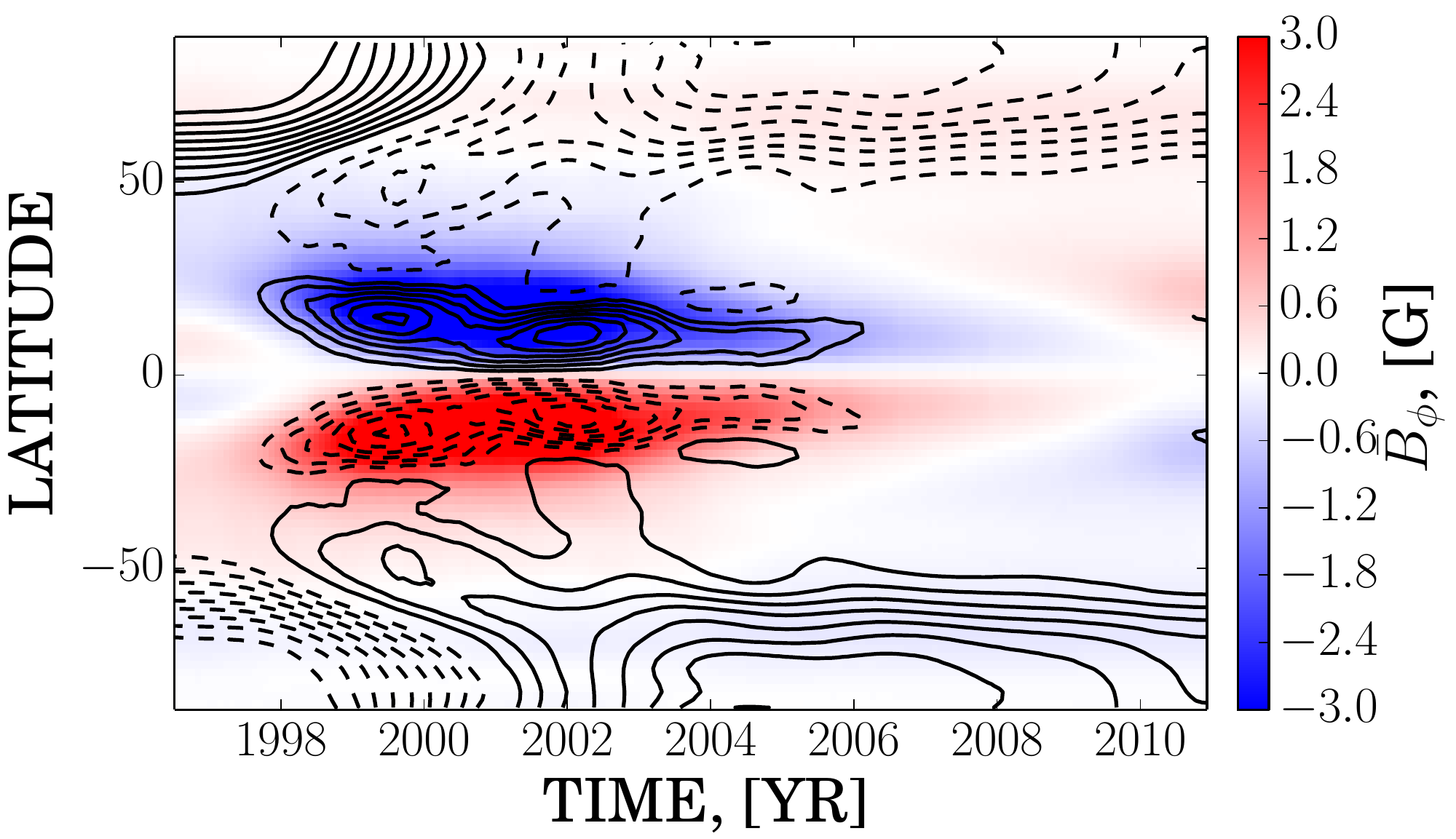}

\caption{\label{fig:The-components-of} Radial ($\bar{B}_{r}$, contours) and toroidal ($\bar{B}_{\phi}$,
background images) components of large-scale magnetic field for
(a) case 1 (odd and even modes) and (b) case 2 (even modes, only)
computed using first 11 modes in the equations (\ref{eq:br},\ref{eq:bf}). Positive (solid) and negative (dashed) contours are drawn at equal steps in amplitude of within the range of $\pm6$G.}
\end{figure}

Figure \ref{fig:The-components-of} shows the components of the large-scale
magnetic field. There, again, we use the first 11 modes in equations
(\ref{eq:br},\ref{eq:bf}). The obtained evolution of $\bar{{B}}_{r}$
is in agreement with results of \citet{ul13}. The pattern of $\bar{{B}}_{\phi}$
in Figure \ref{fig:Components-of-the}b (even modes) closely
resembles Figure \ref{data-smooth}a. We also find that the phase
relation $\bar{B}_{r}\bar{B}_{\phi}<0$ holds in equatorial region
\citep[see also, ][]{stix76a,yosh76}. \cite{2003AdSpR..32.1835B}
argued that this relation is tightly related with the sign of the
magnetic helicity density. Figure \ref{fig:mhl} supports this conjecture
for the asymmetric (relative to solar equator) part of the magnetic
helicity density. 
\begin{figure}
a)\includegraphics[width=0.47\columnwidth]{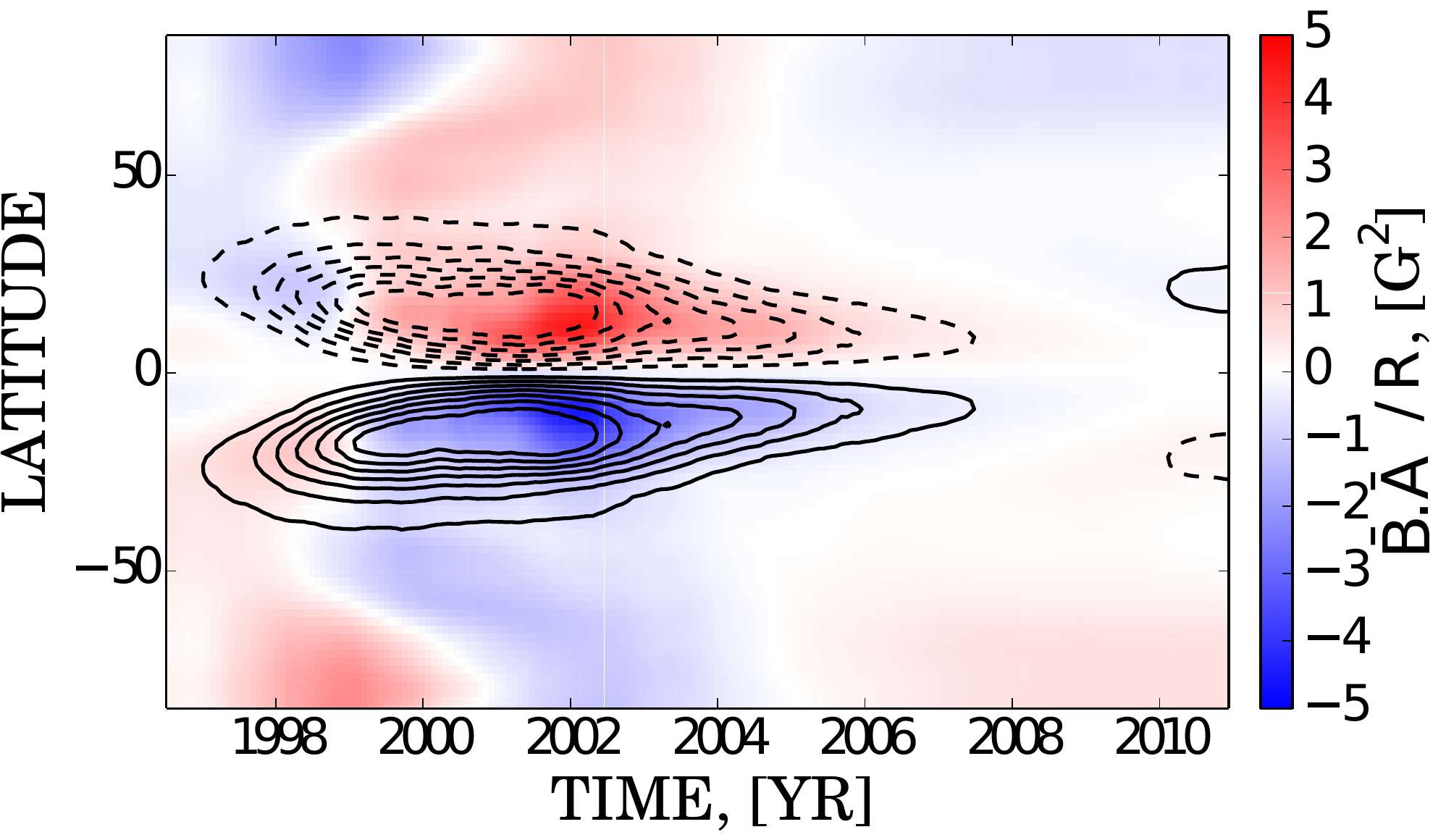}b)\includegraphics[width=0.47\columnwidth]{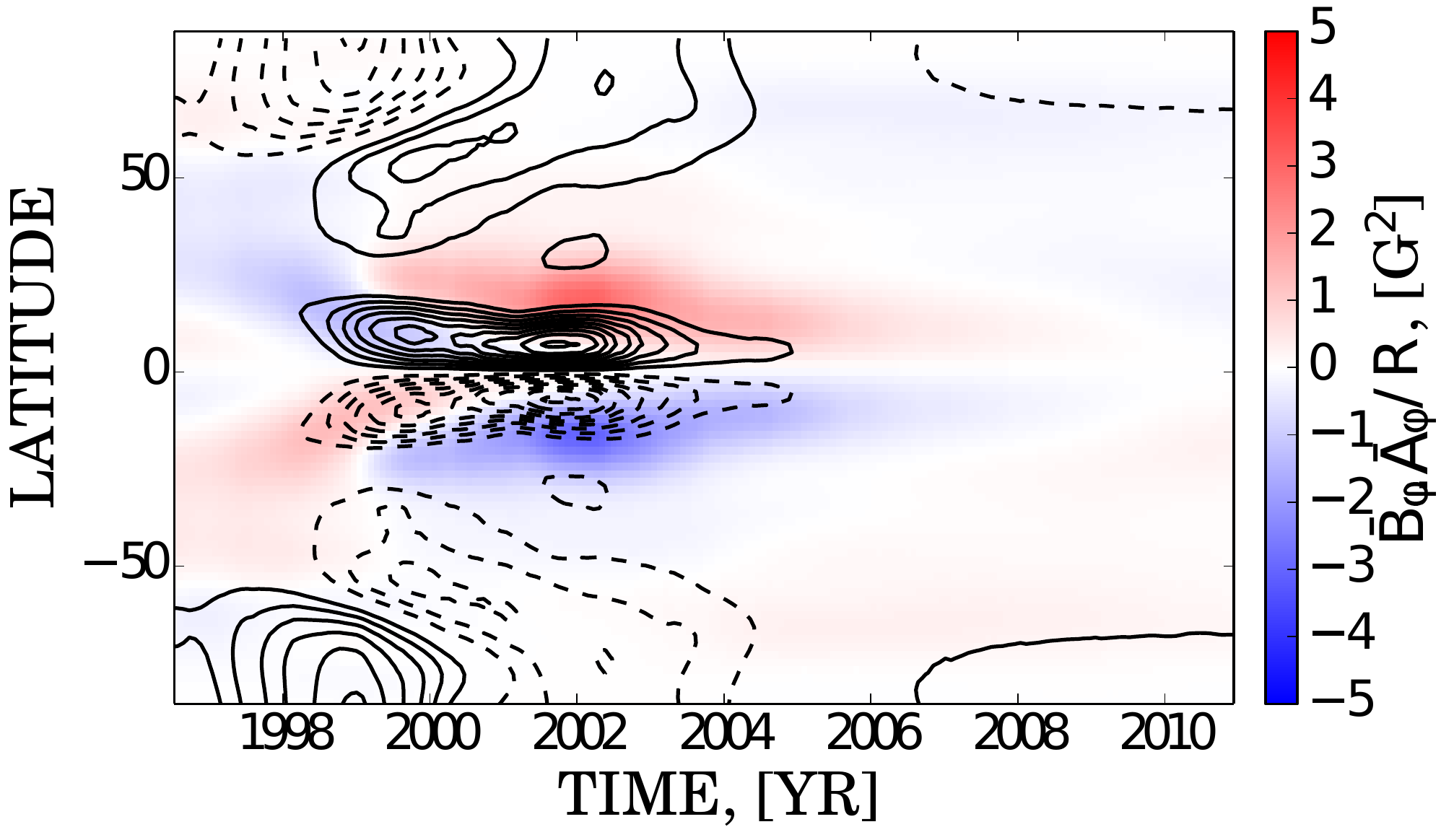}

\caption{\label{fig:mhl}The magnetic helicity density for the asymmetric (relative
to solar equator) part of the large-scale magnetic field. The panel
a) shows the $\bar{\boldsymbol{\mathbf{A}}}\cdot\bar{\mathbf{B}}$
(background images) and the toroidal magnetic field (contours {are
within the range of $\pm3$G}); b) for the same case, the $\bar{A}_{\phi}\bar{B}_{\phi}/R$
(background images) and the $\bar{A}_{r}\bar{B}_{r}/R$ (contours
{are within the range of $\pm5$G}), both values vary within
the same range of magnitude.}
\end{figure}

Having the radial and toroidal components of magnetic field and vector
potential, we are now able to compute the corresponding contributions
to  magnetic helicity density 
$\bar{\boldsymbol{\mathbf{A}}}\cdot\bar{\mathbf{B}}=\bar{A}_{\phi}\bar{B}_{\phi}+\bar{A}_{r}\bar{B}_{r}$.
 The distribution of magnetic helicity density
in cycle 23 (Figure \ref{fig:mhl}a) shows a strong hemispheric asymmetry,
with positive/negative helicity in the northern/southern hemispheres.
This hemispheric asymmetry is opposite in sign to the hemispheric
helicity rule found in active regions. There is no contradiction here.
In the dynamo theory, the active regions are thought to represent
the ``small-scale'' magnetic fields \citep[see][for further discussion]{2003AdSpR..32.1835B},
while in this paper we derive helicity of large-scale fields (in mean-field
dynamo terminology). The fact that large-scale helicity derived by
us has an opposite sign to helicity of active regions is in agreement
with the notion that the dynamo produces helicity of two opposite
signs segregated by their spatial scales.

Patterns of toroidal ($\bar{A}_{\phi}\bar{B}_{\phi}$) and
radial ($\bar{A}_{r}\bar{B}_{r}$) components of magnetic
helicity density are quite different (Figure \ref{fig:mhl}b).
{ Comparing Figure \ref{fig:mhl} and Figure \ref{fig:tot}a we conclude that
the total helicity in polar regions is defined by
$\bar{A}_{r}\bar{B}_{r}$ contribution. In equatorial regions, both
$\bar{A}_{r}\bar{B}_{r}$ and  $\bar{A}_{\phi}\bar{B}_{\phi}$ have the
same sign.  Despite the difference in spatial distributions of $\bar{A}_{r}\bar{B}_{r}$ and
$\bar{A}_{\phi}\bar{B}_{\phi}$,} their total surface integrals are about
equal ($\int\bar{A}_{\phi}\bar{B}_{\phi}d\mu=\int\bar{A}_{r}\bar{B}_{r}d\mu$),
as verified by direct computations using our data (for a mathematical derivation
of this equality, see Appendix). Therefore, one can compute the total surface helicity, 
$\mathcal{H}_{S}=\int\bar{\boldsymbol{\mathbf{A}}}\cdot\bar{\mathbf{B}}d\mu$,
using only one of two parts as suggested by \citet{2003AdSpR..32.1835B}.
The estimation of $\mathcal{H}_{S}$ separately for the northern and
southern hemispheres still requires both $\bar{A}_{r}$
and $\bar{A}_{\phi}$.

\begin{figure}
a)\includegraphics[width=0.3\columnwidth]{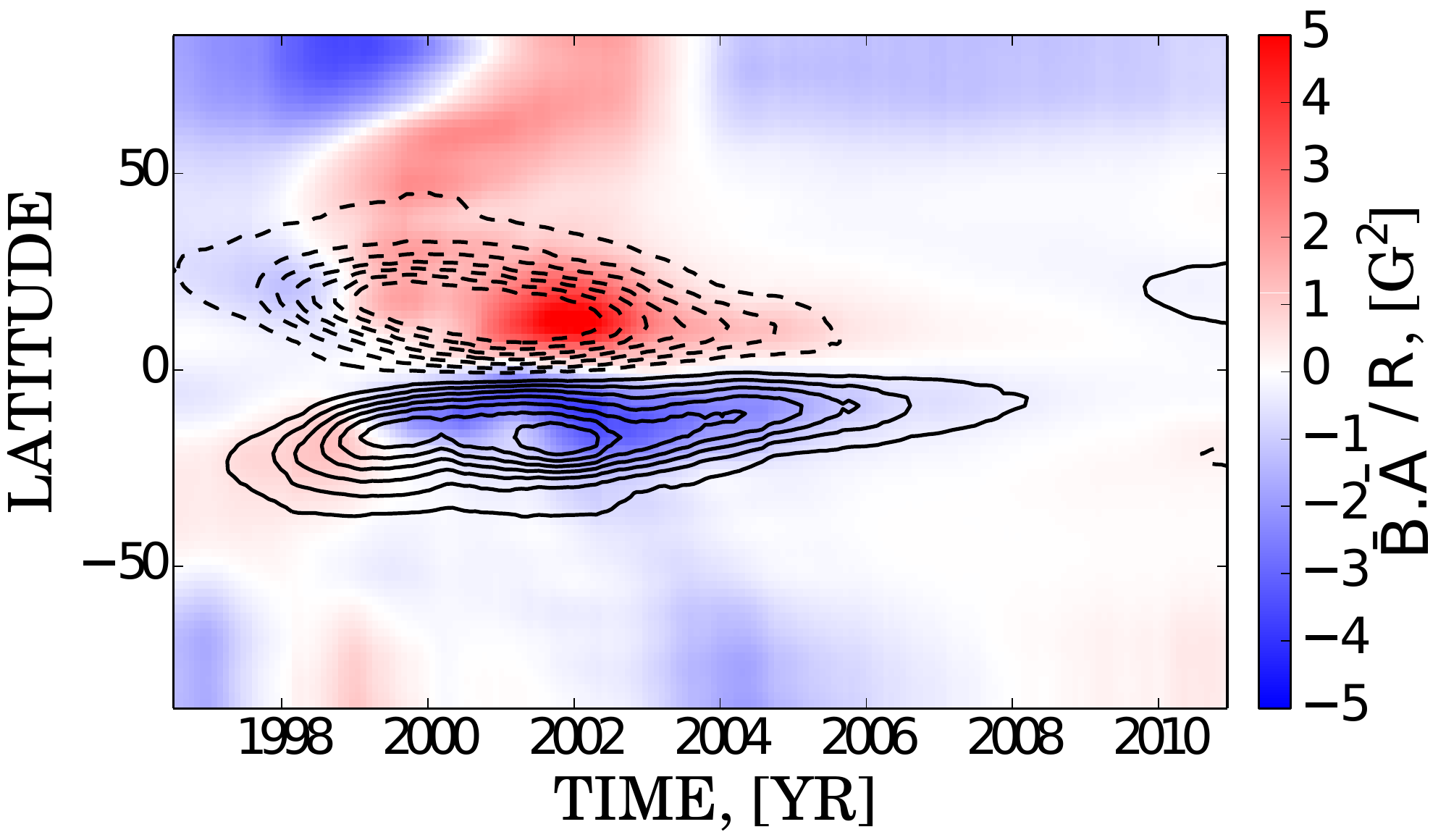} b)\includegraphics[width=0.3\columnwidth]{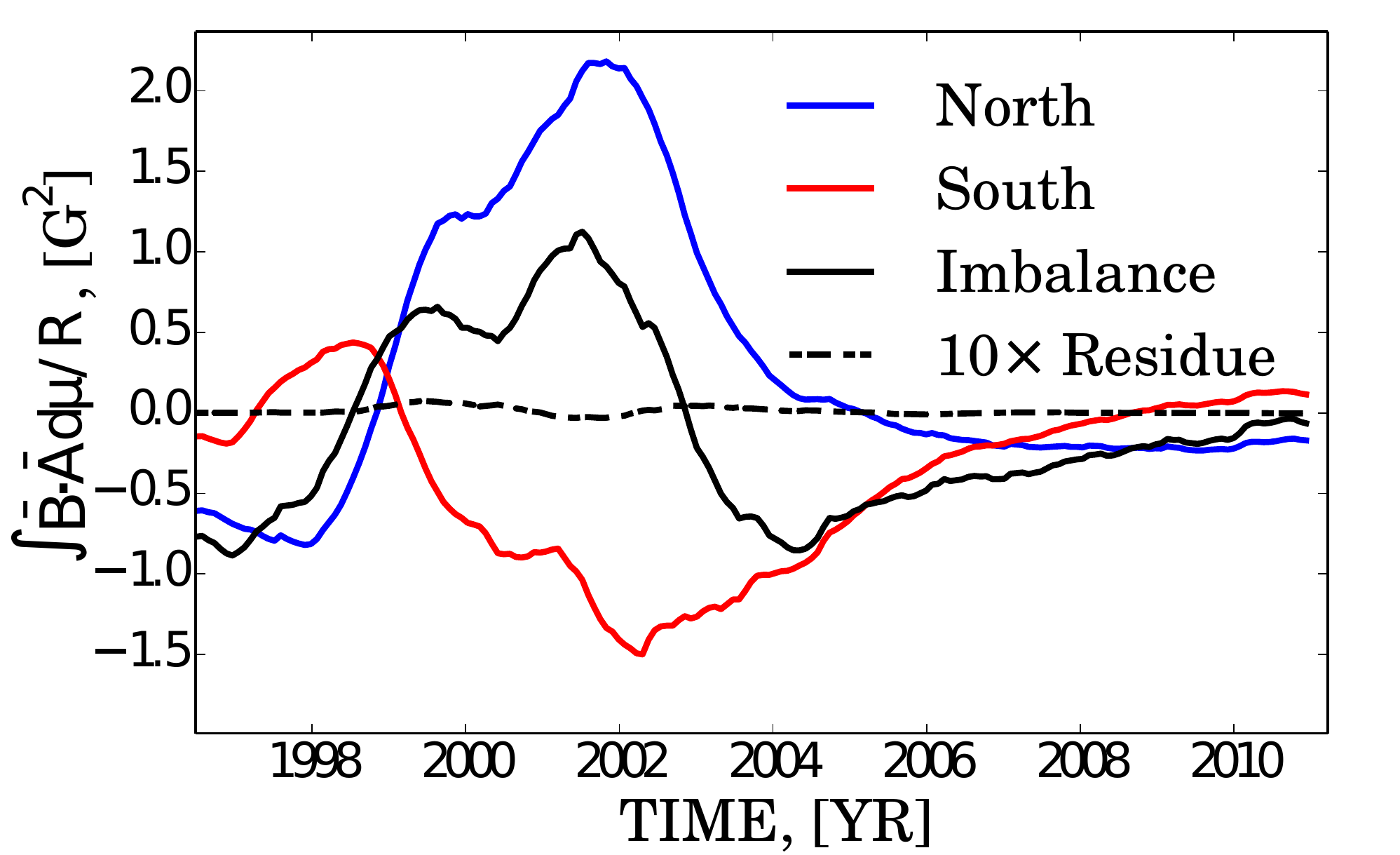}
c)\includegraphics[width=0.3\columnwidth]{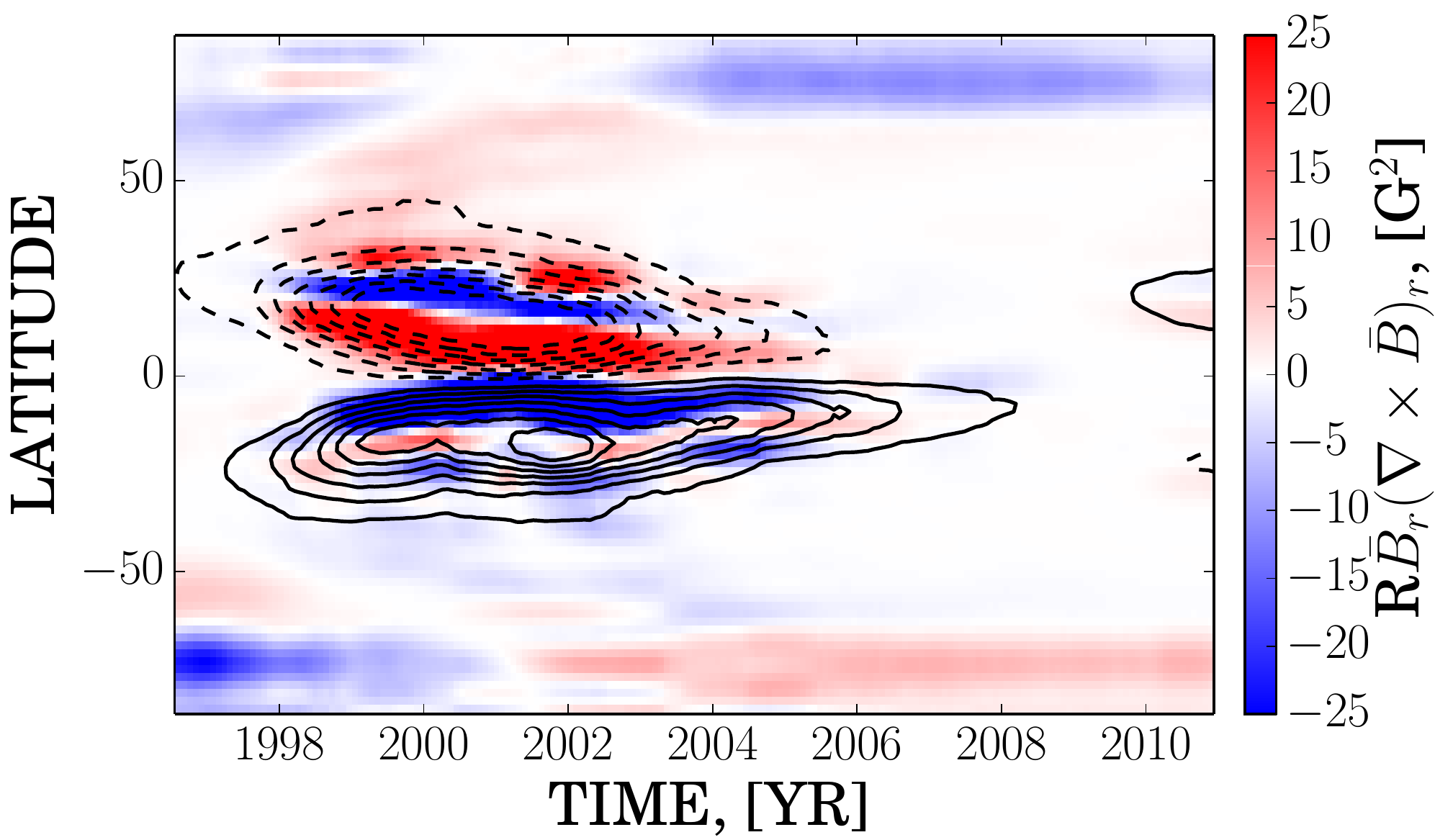}

\caption{\label{fig:tot}Magnetic helicity density (background
image) and the large-scale toroidal field (contours, panel (a),
and magnetic helicity density integrated over each hemisphere (panel b). 
Contribution from the modes with number larger than
11 is shown as ``residue''. Panel (c) show current helicity
density (background image) and the large-scale toroidal field (contours).
Contours are drawn for the same levels as in Figure \ref{fig:mhl}.}
\end{figure}

\begin{figure}
a)\includegraphics[width=0.47\columnwidth]{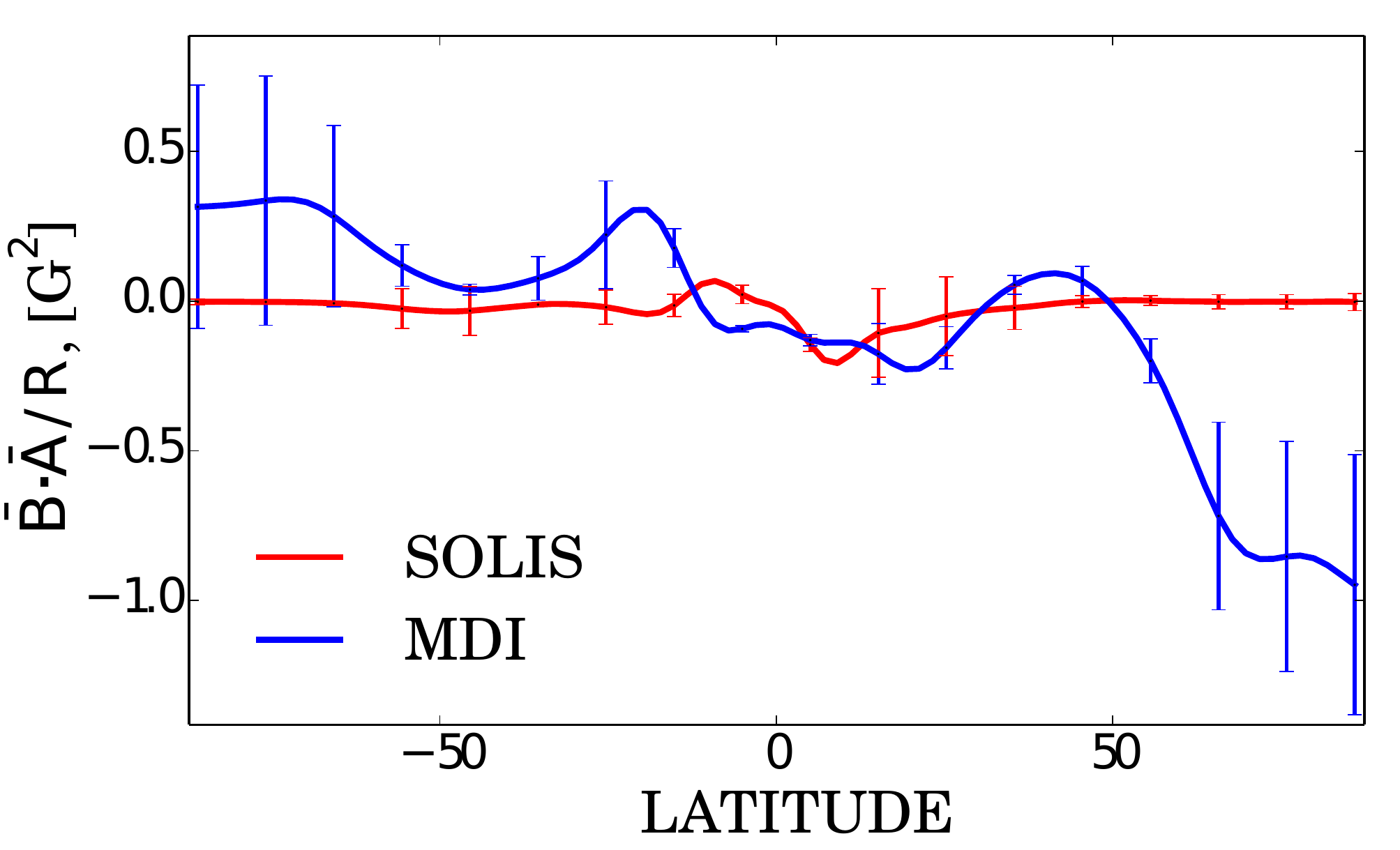}b)\includegraphics[width=0.47\columnwidth]{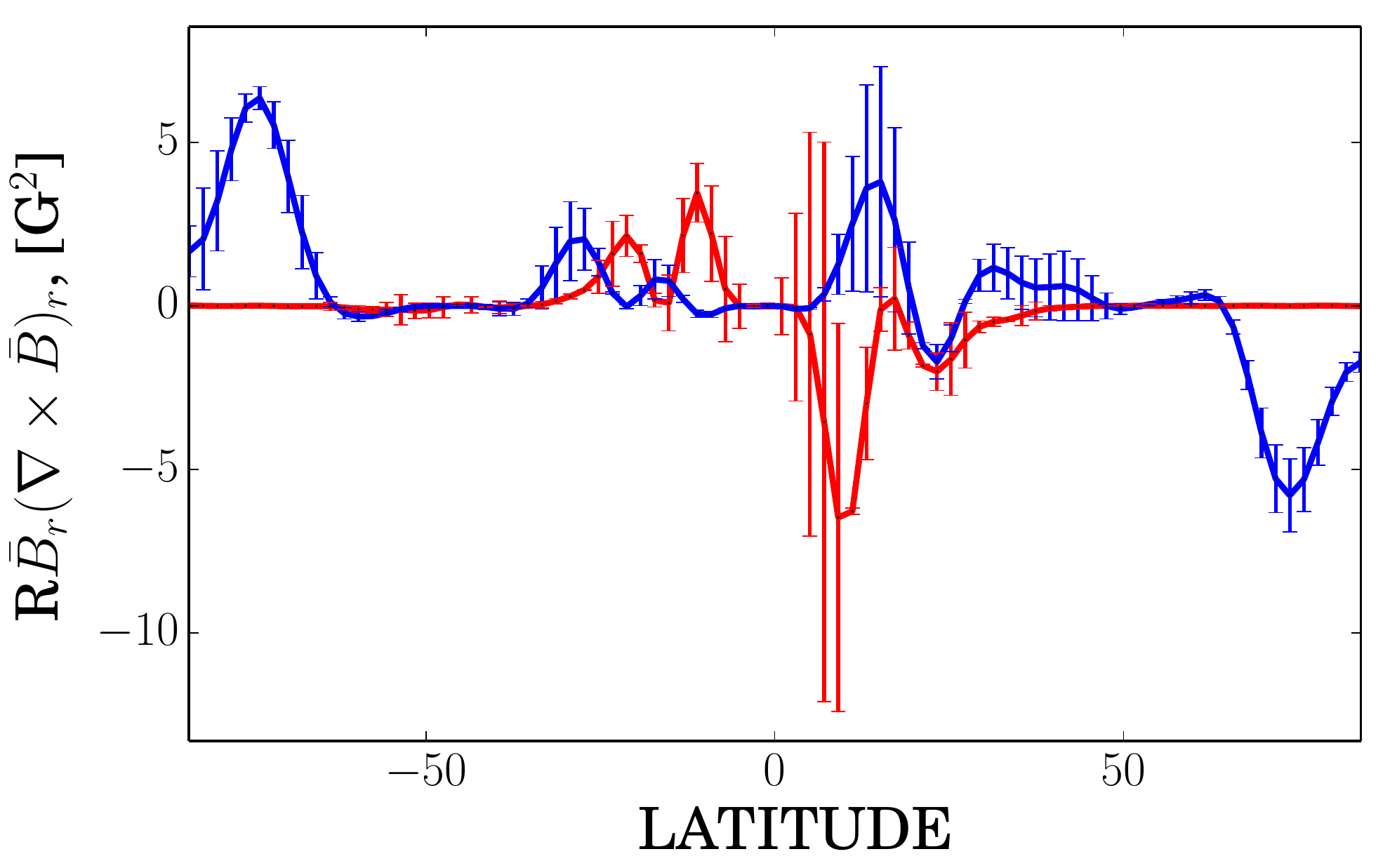}

\caption{\label{fig:tot-1}Latitudinal profiles of magnetic helicity
density (a) and current helicity density (b) for
SOLIS and MDI data sets. The 90 \% confidence intervals are shown as error bars.}
\end{figure}

Modern measurements of solar vector magnetic fields are normally restricted
to a single layer in solar atmosphere (typically, the photosphere).
These observations are insufficient to derive the true magnetic helicity.
Instead, various proxies of helicity are used. 
Figure \ref{fig:tot}c shows evolution of one of these helicity proxies, the radial 
component of current helicity density, $\bar{B}_{r}\left(\nabla\times\bar{\mathbf{B}}\right)_{r}$.
In comparison with true magnetic helicity density (Figure \ref{fig:tot}a),
$\bar{B}_{r}\left(\nabla\times\bar{\mathbf{B}}\right)_{r}$
shows a more complex pattern. While
on average the current helicity density follows the same hemispheric
sign-asymmetry as the magnetic helicity density, during the maximum of solar cycle 23 
the $\bar{B}_{r}\left(\nabla\times\bar{\mathbf{B}}\right)_{r}$
exhibits a distinct ``zebra'' pattern with opposite helicity bands present
in both hemispheres. Whatever these bands persist through the minimum
of cycle 23 is not clear as our data
are insufficient to make a definite conclusion about this. Similar
``zebra'' patterns in the $\bar{B}_{r}\left(\nabla\times\bar{\mathbf{B}}\right)_{r}$
were found in the past \citep[e.g.,][]{2000ApJ...528..999P,pev03,2013ApJ...772...52G}.
\citet{pev03} speculated about a possible relation between the latitudinal
bands of current helicity density and the subphotospheric pattern
of torsional oscillations.

Figure \ref{fig:tot-1} compares latitudinal profiles of 
$\bar{\boldsymbol{A}}\cdot\bar{\mathbf{B}}$
and $\bar{B}_{r}\left(\nabla\times\bar{\mathbf{B}}\right)_{r}$
computed using data from SOHO/MDI (about year 2011) and SOLIS/VSM
(Figure \ref{nso1}c, year 2012). 
The 90 \% confidence
interval was computed in the same manner as for data shown in 
Figures \ref{fig:LOS}c and \ref{nso1}c. The residual contribution of modes higher
that n=11 (equations(\ref{eq:aa}--\ref{eq:arf})) is an order of magnitude
smaller then the contribution of first 11 modes.
While both 
magnetic helicity density and $\bar{B}_{r}\left(\nabla\times\bar{\mathbf{B}}\right)_{r}$
exhibit the hemispheric helicity rule in both datasets, there are some differences. 
For example, the latitudinal profiles of helicity can differ because of 
evolutionary changes (MDI and VSM data included in this comparison correspond to periods, 
which are about one year apart). 
Other sources of difference could include difference in sensitivity to 
magnetic field
and the noise levels, as well as the treatment of polar fields. 

\section{Discussion and Conclusions}

Using synoptic charts from SOHO/MDI, we, for
the first time, reconstruct the magnetic helicity density of global 
axisymmetric field of the Sun. In solar
cycle 23 the global axisymmetric magnetic field exhibits positive magnetic
helicity in the northern hemisphere, and negative one in the southern.
In general, such reconstructions require a knowledge of the $\bar{B}_{r}$
and $\bar{B}_{\phi}$ components of the axisymmetric magnetic field.
In the past, vector global magnetic field components were reconstructed
via various approaches \citep{2000ApJ...528..999P,2005ApJ...620L.123U,hoeks10,2012SoPh..280..379M}.
Here we used synoptic charts of LOS magnetic field corresponding to different 
longitudinal offsets relative to central meridian to compute $\bar{B}_{r}$ and $\bar{B}_\phi$, see equation (\ref{eq:torf}). Derived $\bar{B}_{r}$ agrees well with
the $\bar{B}_{r}$ provided by the MDI team, which we see as indirect validation 
of our method.

Based on the analysis of dynamo equations, \citet{2003AdSpR..32.1835B} 
suggested that the magnetic helicity of global magnetic field
in solar cycle 23
should be positive/negative in the northern/southern hemisphere.
\citet{pip13M} analyzed the distributions of magnetic helicity for
large- and small-scale magnetic fields in the axisymmetric mean-field
dynamo taking into account the conservation of the total magnetic helicity
in the dynamo processes. They concluded (see their Figures 2e,d and 5)
that magnetic helicity density of large-scale field should have
positive sign in the northern hemisphere and negative sign in
the southern hemisphere during the most part of the magnetic cycle. 
Our present results provide observational support to these early theoretical predictions. 
We find that during most of cycle 23, global magnetic fields
exhibited a persistent pattern of positive/negative helicity in the northern/southern
hemispheres. 

In respect to helicity of active region magnetic fields
(small-scale in the framework of this discussion), the hemispheric
helicity rule is negative/positive in the northern/southern hemispheres
\citep[][and references therein]{see1990SoPh,pev95,zetal10}. Taken
together, these two results support the notion that the solar dynamo
creates helicity of two opposite signs as was suggested in early papers.
However, helicity of both signs seem to cross the solar photosphere.
We further found that the hemispheric helicity rule for global magnetic
fields exhibits sign-reversals in early and late phases of cycle 23.
If the helicities of small- and large-scale fields are tied together,
this should imply a need for similar reversals in the hemispheric helicity rule for
active regions. Alas, while some researchers claimed observing reversals
in the hemispheric helicity rule near the minimum of solar cycle 22
and 23 \citep{bao2000,hag05}, others were not able to find them 
\citep{pev2001,pev2008,2013ApJ...772...52G}.
Clearly, this question about possible reversals of the hemispheric
helicity rule for active region magnetic fields needs to be re-examined.
Although, the predictions of the model by \cite{pip13M}
agree qualitatively with the results reported in our paper,  there are
some differences related to the shape of the helicity density
patterns. For example, our present results suggest that magnetic helicity
density pattern of the same sign can extend from the equator to the
poles which is not seen in the model.  Similarly, the pattern of 
$\bar{\mathbf{A}}\cdot\bar{\mathbf{B}}$ of reverse sign penetrates to equatorial
regions during the minima of cycle around year 1997 and 2009.
If the reversals of the hemispheric helicity rule are real, this will
pose a challenge for some proposed mechanisms of helicity generation
\citep[e.g., helicity  generation by the differential
rotation,][]{2000JGR...10510481B}.

{ As an alternative explanation, our results could be interpreted
  in the framework of helicity of axisymmetric and non-axisymmetric
  parts of global magnetic fields. 
In that model, the axisymmetric component of helicity (derived in this
article) follows the hemispheric helicity rule of opposite in sign to
the non-axisymmetric component (associated with active regions).  Such
a possibility was raised by Zhang (2006), who used Berger \& Ruzmaikin
(2000) data to show that helicity flux of non-axisymmetric modes 
was opposite in sign to helicity of axisymmetric (m=0) mode.}

{ One may also question the importance of vector magnetic field measurements 
for studies of global helicity, when the line-of-sight data seem to provide reasonable results.
Here we presented the first ever derivations of magnetic
helicity density of global field based on vector synoptic maps.
While we see similarities in distribution of global helicity derived from
LOS and vector data, there are also some differences. 
For example, synoptic maps of toroidal field derived from LOS data show more
or less uniform distribution of the magnetic field polarity of one
sign suggested by the the Hale polarity law. In addition to that pattern,
vector field maps show mix of two different polarities: one is more concentrated and 
other is somewhat diffused (Figure
3b).  The diffuse component (of toroidal field) seems to correspond to a trailing polarity
field. Such component of large-scale field is not present in the maps of toroidal flux 
derived from the LOS  magnetic fields. The vector field data are limited to
cycle 24, and thus, are insufficient to conclude if there are any changes in this 
diffuse component of toroidal field with solar cycle. This and other differences
between derivations based on LOS or vector field data 
require further investigation.}

Our findings indicate that helicity of the large-scale magnetic fields
is imbalanced between the northern and the southern hemispheres in
different phases of solar cycle. However, when taken over the entire
cycle, the positive and negative helicity of large-scale magnetic
field is well-balanced. Indirectly, this is in agreement with \citet{geor2009},
who found that helicity injection through the solar photosphere
associated with active region magnetic fields is well-balanced over
the solar cycle 23. On the other hand, \citet{2012ApJ...758...61Y}
reported significant imbalance between helicity fluxes of northern
and southern hemispheres. Our findings (Figure 8c) allow to reconcile 
\citet{geor2009} and \citet{2012ApJ...758...61Y} conclusions.

{Due to limitations of the existing datasets, the observational studies of 
helicity often refer to proxies of current helicity density. It is  usually assumed 
that these proxies represent magnetic helicity 
sufficiently well. Contrary to that, we find that while the general tendencies 
are similar in magnetic and current helicity densities, there are differences, for
example, in small-scale patterns, which may be present in one helicity proxy but are 
absent in the other. For example,} proxy of current helicity, 
$\bar{B}_{r}\left(\nabla\times\bar{\mathbf{B}}\right)_{r}$,
exhibits a distinct ``zebra'' pattern, but no such pattern is present in the 
distribution of magnetic helicity. Early, \citet{2000ApJ...528..999P} and
\citet{2013ApJ...772...52G} reported similar pattern in current helicity density 
of large-scale magnetic fields. The pattern could also be expected from spatial
structure of the dynamo wave of the large-scale magnetic field components
$\bar{B}_{r}$ and $\bar{B}_{\phi}$, which are illustrated in Figure \ref{fig:Components-of-the}a.
We note that in the equatorial regions the inequality $\bar{B}_{r}\bar{B}_{\phi}<0$
holds for the most part of the sunspot cycle (see, Figure \ref{fig:Components-of-the}a).
We also found that modes $b_{r}^{(3)}$ and $b_{\phi}^{(2)}$ dominate,
which means that $\bar{B_{r}}\left(\boldsymbol{\nabla}\times\mathbf{\bar{B}}\right)_{r}\sim b_{\phi}^{(2)}b_{r}^{(3)}P_{2}P_{3}$,
here the sign of the $b_{\phi}^{(2)}b_{r}^{(3)}$ defines the hemispheric
sign rule and the product $P_{2}P_{3}$ defines that zebra pattern as
illustrated in Figure \ref{fig:tot}c and Figure \ref{fig:tot-1}. Thus,
the results shown in Figure \ref{fig:tot}c are expected for any dynamo
model that qualitatively reproduces Figure \ref{fig:Components-of-the}a.

Finally, keeping in mind the approximations which were used in the reconstruction
of components of the global magnetic field of the Sun, our results should be 
considered as preliminary.
Further development in this direction is likely to shed more light on
the role of magnetic helicity in global solar and astrophysical
dynamos.

\acknowledgements{ }

This work utilizes SOLIS data obtained by the NSO Integrated Synoptic
Program (NISP), managed by the National Solar Observatory, which is
operated by the Association of Universities for Research in Astronomy
(AURA), Inc. under a cooperative agreement with the National Science
Foundation. SOHO is a project of international cooperation between
ESA and NASA. We thank Mei Zhang for providing results of her derivations
for comparison with our results. This work has benefited from fruitful
discussions at 2013 Helicity Thinkshop in Beijing, China. Useful discussions
with K.M.Kuzanyan are acknowledged, as well. We thank Axel Brandenburg
for critical reading of manuscript. VVP thanks the National Astronomical
Observatories of China for support of his visits to Beijing, the support
the RFBR grant, 13-02-91158-GFEN-a and the support of the project
II.16.3.1 under the Program of Fundamental Research of SB RAS. AAP
acknowledges partial support from NSF/SHINE Award No. 1062054.


\section{Appendix A}

\setcounter{equation}{0}

\global\long\def\theequation{A\arabic{equation}}

The components of large-scale axisymmetric field and the components
of its vector potential are related with equations:
\begin{eqnarray}
\bar{B}_{r} & = & -\frac{1}{r}\frac{\partial\left(\sin\theta\bar{A}_{\phi}\right)}{\partial\mu},\\
\bar{B}_{\phi} & = & \frac{\sin\theta}{r}\frac{\partial\bar{A}_{r}}{\partial\mu},
\end{eqnarray}
where $\mu=\cos\theta$. Then, using integration by part, we obtain:
\begin{eqnarray}
\int_{-1}^{1}\bar{A}_{\phi}\bar{B}_{\phi}d\mu & = & \int_{-1}^{1}\frac{\sin\theta\bar{A}_{\phi}}{r}\frac{\partial\bar{A}_{r}}{\partial\mu}d\mu\\
 & = & \left.\frac{\sin\theta\bar{A}_{\phi}\bar{A}_{r}}{r}\right|_{0}^{\pi}-\int_{-1}^{1}\frac{\bar{A}_{r}}{r}\frac{\partial\sin\theta\bar{A}_{\phi}}{\partial\mu}d\mu\nonumber \\
 & = & \int_{-1}^{1}\bar{A}_{r}\bar{B}_{r}d\mu.\nonumber 
\end{eqnarray}

\end{document}